\documentclass[11pt]{article}

\usepackage{latexsym}
\usepackage{amsthm}
\usepackage{amsmath}

\newfont{\sdbl}{msbm9}
\newfont{\dbl}{msbm10 at 12pt}
\theoremstyle{definition}
\newcommand{\dz}{{\mbox{\dbl Z}}}
\newcommand{\dr}{{\mbox{\dbl R}}}
\newcommand{\dn}{{\mbox{\dbl N}}}

\newcommand{\sdz}{{\mbox{\sdbl Z}}}
\newcommand{\dc}{{\mbox{\dbl C}}}

\newcommand{\ord}{\mathop{\rm ord}\nolimits}
\newcommand{\bfu}{\mathop{\rm \bf u}\nolimits}
\newcommand{\bfs}{\mathop{\rm \bf s}\nolimits}
\newcommand\arrone{\overset{\partial_1}\longrightarrow}
\newcommand\arrtwo{\overset{\partial_2}\longrightarrow}
\newcommand\indeks{\overset{i-j+k=\alpha}{j\ge 1, i,k\le N}}
\newcommand\indeksa{\overset{l+n=\alpha}{l,n\le N}}
\makeatother
\newtheorem{defin}{Definition}
\theoremstyle{plain}
\newtheorem{prop}{Proposition}

\newtheorem{theo}{Theorem}
\newtheorem{lemma}{Lemma}
\newtheorem{corol}{Corollary}

\newcommand{\Nu}{{\cal V}}
\begin{document}

\centerline{\LARGE \bf Two dimensional KP systems and their solvability}

\bigskip

\centerline{{\Large Zheglov A.B.}
\footnote{Supported by the DFG Schwerpunkt "Globale Methoden in der Komplexen Geometrie"}
\footnote{e-mail address: azheglov@mathematik.hu-berlin.de}}

\bigskip

\centerline{\bf Abstract}

\begin{quote} 
\small{In this paper we introduce new various generalizations of the classical Kadomtsev-Petviashvili hierarchy in the case of operators in several variables. These generalizations are the candidates for systems that should play the role, analogous to the role of the KP hierarchy in the classical KP theory, in a generalized KP theory. In particular, they should describe flows of some generalized geometric data, including those described in \cite{Pa3}, for certain initial conditions. 
The unique solvability of the initial value problem for the generalized KP hierarchies is established. 
The connection of these systems with universal families of isospectral deformations of certain pairs of commuting differential operators is opened. To prove the solvability of the systems we generalize several results from the works of M.Mulase (\cite{Mu}) and A.N.Parshin (\cite{Pa}).} 
\end{quote}

\section{Introduction}

In \cite{Pa} A.N. Parshin offered a generalization of  the classical KP-hierarchy and 
studied different properties of this system: the conversations laws, Zaharov-Shabat 
equations and some others. The generalized KP-hierarchy was there interpreted as a dynamical system on some infinite-dimensional variety.  
Recall that the classical KP-hierarchy has the following 
Lax form:
\begin{equation}  \label{KP}
\frac{\partial L}{ \partial t_n}
=
\left[ (L^n)_+, L)       \right] ,  \mbox{\quad}  n \ge 1 \mbox{,}
\end{equation}
where $L = \partial + u_{-1} \partial^{-1} +  \ldots \in P$ is  a pseudodifferential 
operator,  $u_i \in k ((x))[\log(x)][[t_k]]$
and $P = P_+ \oplus P_-$, where $P=k((x))((\partial^{-1}))$ is the ring of 
pseudodifferential operators (in one variable), $P_+$ is the subring of differential 
operators. 

The classical KP-hierarchy is only a starting point of a huge KP theory developed since 1970s or even earlier, which have, beyond other, a rich algebraic structure. Under the algebraic structure we mean here a so-called Krichever correspondence, which describes correspondences between certain solutions of the classical KP (KdV, etc.) equations and hierarchies, certain geometric datas (which  consist of an algebraic complete curve, a point, a torsion free sheaf and a trivializations of it in the classical case), rings of commuting ordinary or matrix differential operators, points and moduli varieties  in a universal grassmanian, $\theta$-functions of jacobians of curves and $\tau$-functions (for detailed explanation see, for example, the review \cite{Mu3} and other references cited there).  
A generalization of the KP-hierarchy should play a role of a system, which describe, for certain initial data, flows of some generalized geometric data, which should include algebraic varieties of higher dimension. In works \cite{Pa3}, \cite{O} the so-called Krichever map was generalized. In the classical case this is a map that sends the geometric Krichever data to a subspace of a one-dimensional local field, which can be interpreted as a point of an infinite-dimensional grassmanian. 

One of the important steps in generalizing the classical KP-theory is studying the  solutions of  generalized KP-hierarchies and their connections  with generalized geometric datas. 
In \cite{Mu} M.Mulase solved the  Cauchy problem for the classical and more complicated 
KP-hierarchies. Parshin's generalization deals with the ring of pseudodifferential 
operators in $n$ variables. Except the properties proved in \cite{Pa} there remain 
unsolved a lot of questions, in particular, it was not clear if the Cauchy problem has a solution in this 
case, is it true that the Parshin system is a master equation of all isospectral 
deformations of some differential operators in 2 variables, is there a  connection 
with the problem of classification of all commutative subrings in the ring of 
differential operators in several variables, are there some geometrical solutions of 
these systems and so on.   

In \cite{Zh} a number of these questions has been solved. In particular, we proved that the Parshin system itself has no nontrivial solutions, but certain its subsystems do have. The subsystems are parametrized by linear functions $\alpha :\dz_+\rightarrow \dr_+$ and satisfy the same properties as the original one. Moreover, they are uniquely solvable in a certain ring of time-dependend pseudo-differential operators. 

In this paper we continue to study the new systems. We show that 
the original Parshin system (which arise naturally in the framework of the theory of higher dimensional local (skew)-fields) is 
a "half" of a master equation of all universal families of isospectral deformations for 
any pair of normalized commuting differential operators that have some additional condition on their orders (conjecturally for pairs of completely integrable operators in the sense of \cite{Et}, \cite{Ga}). Since the pair of 
normalized commuting differential operators can be choosen arbitrarily, we come to the necessity of reproving the results from \cite{Zh} under weaker assumptions. In particular, we prove that the Parshin system has  
no nontrivial solutions even on a bigger variety than in \cite{Pa}, \cite{Zh}. This means that the bigger system, the master equation of isospectral deformations, has the same property. 

From the other hand side, one can consider subsystems of the master equation and try to solve them. In this paper we call such subsystems {\it modified Parshin's hierarchies}. All these systems satisfy 
the same properties as the original Parshin system. They can be interpreted as master equations of all universal families of isospectral deformations of some pairs of commuting $\alpha$-differential operators (the definition of this notion is given in section 6). 

We study them in a sufficiently general situation of a ring of pseudodifferential operators over a (commutative) ring $A$, which satisfy certain properties listed in section 3.1. In particular, $A$ can be equal to the ring $k[[x_1,x_2]]$ or $k((x_1))((x_2))[\log x_1,\log x_2]$. 
We show that all such systems parametrized by functions $\alpha :\dz_+ \rightarrow \dr$ such that $\alpha (0)\le 0$ (and denoted by $(KP)_{\alpha}$) are uniquely solvable in a certain generalized ring of time-dependend pseudo-differential operators for all initial values. For the constant function $\alpha =0$ the system 
$(KP)_{\alpha}$ covers the classical KP-hierarchy, and, if we assume that $\alpha$ can have the value "$\infty$", then for the constant function $\alpha =\infty$ the system $(KP)_{\alpha}$ is the master equation described above. 

The whole master equation gives a necessary condition on a time-dependness of the ring of coefficients of operators. Recall that in one-dimensional situation this ring is obtained as a completion of the ring of polynomials in infinite many times with respect to a discrete valuation (see \cite{Mu}). In two-dimensional case the ring is defined as a completion of the ring of polynomials in infinite many times with respect to a topology, whose model comes from the topology on a two-dimensional local field (see \cite{PaFi} or \cite{Fe} for background on the theory of local fields). Such a topology is a weakest one, for which the unique solvability of the modified KP-systems is established (see prop. \ref{SW}).

To solve the Cauchy problem for the modified KP-systems we generalize the classical method. Namely, we first prove the equivalence of the modified KP-systems and obviously modified Sato-Wilson systems (in \cite{Mu} the Cauchy problem was solved exactly for Sato-Wilson systems, in the case of commutative operator's coefficients they are equivalent to KP-systems, see the discussion after lemma 1.3 there). The commutativity of the ring $A$ is important on this step (see sections 3.2 and 4.1).

Then we find a solution of the Sato-Wilson systems using a generalization of the {\it Birkhoff decomposition}. The original Birkhoff decomposition gives a factorization of a loop group into a product of subgroups of loops of special form (\cite{PS}); it was then generalized in \cite{Mu}, where the loop groups were replaced by groups of infinite order micro-differential  operators. The last groups were defined as groups of certain invertible elements in a ring of extended time-dependent pseudo-differential operators. These operators can be represented as infinite series with certain valuation growth condition on their coefficients. We generalize the Birkhoff decomposition of \cite{Mu} considering groups of certain invertible elements in a ring of extended time-dependent {\it iterated} pseudo-differential operators. These operators can be represented as iterated infinite series with certain more complicated valuation growth conditions on their coefficients. Now a solution of a system $(KP)_{\alpha}$ can be obtained from a solution $U$ of the universal equation $dU=\omega_{\alpha}^{N_0}U$, where $\omega_{\alpha}^{N_0}$ is defined by formula (\ref{uni}) below, in the same way as in \cite{Mu} (see section 4.2 below). Notably, the solution of the Sato-Wilson system exists also in the case of a non-commutative ring $A$, like in  \cite{Mu}.

We would like to emphasize that the modified Sato-Wilson systems can not be solved just by reducing them to systems considered in \cite{Mu}, though each modified Sato-Wilson system can be represented, with help of some reordering of indeces, as a system from  \cite{Mu} with non-commutative coefficients. Actually, the main problem of these systems is that it is not clear a priory if there exists a reordering  of indeces that brings a solution of a system from \cite{Mu} to a solution of a modified Sato-Wilson system. So, we should again go through a generalization of the Birkhoff decomposition. 

The modified KP-systems should play the role discussed above for a generalization of the classical KP theory. Nevertheless, we almost don't  explain the connection between the solutions of systems and generalized geometric datas in this paper --- this is a material of another paper. 

\smallskip

Recently there have been made several attempts to generalize some other aspects of the KP theory to higher dimensions developing some ideas appeared in works of Nakayashiki, \cite{N}. These are the works \cite{R} and \cite{Mi}, where, in particular, some new systems that describe flows on Picard varieties (or their extensions) of higher dimensional varieties obtained. These are also the systems of KP type, but for operators with {\it matrix coefficients}. It would be interesting to compare various systems. 

\smallskip 

Here is a brief overview of this paper.

In section 2.1 we recall some definitions, set up the notation 
and give one example. After reading this section the reader can go directly to section 6 to find the motivation of further reseach from the point of view of the isospectral deformation problem described above. 

In section 2.2 we show that the Parshin system, which is the "half" of the master equation from section 6, independently of time-dependentness of operator's coefficients, has only trivial solutions.  

In section 3.1 we introduce a series of modified KP systems, which are subsystems of the master equation, and define the ring of time-dependent operator's coefficients and the ring of extended iterated pseudo-differential operators discussed above.  

In section 3.2 we prove several important technical results. The most important result is theorem \ref{Parshin} that is a generalization of the conjugacy theorem from \cite{Pa}. It says that any pair of commuting generalized time-dependent pseudo-differential operators with main orders equal to (0,1) and (1,0) can be conjugated by one zeroth order invertible operator to a simple canonical form associated to each such pair. 

In section 4.1 we prove that the modified KP-systems are equivalent to a certainly defined modified Sato-Wilson systems. In particular, we prove that the modified KP-systems are equivalent to an analog of Zaharov-Shabat equation on connections. We also classify the set of all admissible operators (lemma \ref{admissible}). This classification occurs to be much more difficult than in one-dimensional case (\cite{Mu2}). 

In section 4.2 we generalize several proofs of Mulase from \cite{Mu} and show the unique solvability of the modified Sato-Wilson systems for arbitrary initial conditions and arbitrary (non-commutative) ring of coefficients $A$. The solutions, nevertheless, may not belong to the ring of usual (not extended) iterated pseudo-differential operators.

In section 5 we give examples of modified KP systems and their initial conditions, whose  solutions  belong to the usual ring of iterated pseudo-differential operators, and show that  the solutions are not trivial in general (that is, the corresponding solutions of the modified Sato-Wilson systems are not admissible). 

In section 6 we generalize the classical definition of a family of isospectral deformations of an ordinary monic differential operator to the case of a pair of monic commuting operators. We then derive, as in the classical case, that the problem of finding of a universal family of isospectral deformations for a pair of such operators (satisfying some additional condition on their order) is equivalent to the problem of finding a solution of an equation that have a Lax form, like a classical KP system. The original Parshin KP-hierarchy is a part of this equation. After that we introduce a notion of $\alpha$-differential operators and show that the modified KP systems are the master equations of all universal families of isospectral deformations of certain pairs of monic commuting $\alpha$-differential operators.

\smallskip

All proofs of this paper are selfcontained and don't depend on the paper \cite{Zh}.

\smallskip

{\it Acknowledgments.} I am very grateful to Professor A.N.Parshin for his permanent attention to this work and for discussions and advises.  I am very grateful to Professor H. Kurke and to D.Osipov for the helpful advises and various discussions.

\section{Generalized  KP-hierarchy }

\subsection{General setting} 
In this paper we use the notation from \cite{Pa}. We will work with the 
following objects: \\
an associative algebra $A$ over a field $k$ of characteristic zero with the unity 1 and with two derivations $(\partial_1, \partial_2)$ such that $\partial_1\partial_2=\partial_2\partial_1$ and $ker(\partial_1)\cap ker(\partial_2)=k$; everywhere, if another is not mentioned, the algebra $A$  is assumed to be commutative;\\
the ring of formal pseudo-differential operators $E=A((\partial_1^{-1}))((\partial_2^{-1}))$. 

Recall that such a ring can be defined iterately, and for any ring $B$ and its derivation $\partial$ the ring $B((\partial^{-1}))$ is defined as a left $B$-module of all formal expressions 
$$
L=\sum_{i>-\infty}^{n}a_i\partial^i, \mbox{\quad} a_i\in B
$$ 
with a multiplication defined according to the Leibnitz rule:
$$
(\sum_ia_i\partial^i)(\sum_jb_j\partial^j)=\sum_{i,j,k\ge 0}C_i^ka_i\partial^k(b_j)\partial^{i+j-k},
$$
where 
$$
C_i^k=\frac{i(i-1)\ldots (i-k+1)}{k(k-1)\ldots 1}, \mbox{\quad if $k>0$, } C_i^0=1.
$$

If $L=\sum_{i\le m}a_i\partial_2^i\in E$ and $a_m\ne 0$, then $m:=\ord_{\partial_2}(L)$ will be called the {\it order} of the operator $L$. The function $\ord_{\partial_2} ( . )$ defines a decreasing filtration $E_{.}:\ldots \subset E_{-1}\subset E_0\subset \ldots $ of vector subspaces $E_i=\{L\in E:\ord _{\partial_2}(L)\le i\}\subset E$ such that $\cap_{i\in\dz}E_i=0$ (i.e. $\ord_{\partial_2}(0)=-\infty$). Analogously, one can define the function $\ord_{\partial_1}(.)$ on the ring 
$A((\partial_1^{-1}))$. Further we will sometimes use the notation $\ord$ instead of $\ord_{\partial_2}$. 

For a given $L=\sum_{i\le m}a_i\partial_2^i\in E$ and $a_m= \sum_{j\le n}a_{mj}\partial_1^j$, $a_{mn}\ne 0$ the element $a_{mn}$ will be called the {\it highest coefficient} of the operator $L$. If $a_{mn}=1$, then $L$ will be called {\it monic}. 

We have the decomposition of $E$ in a direct sum of subspaces 
$$
E=E_{+}+E_{-},
$$
where $E_{-}=\{L\in E:\ord_{\partial_2} (L)<0\}$ and $E_{+}$ consists of the operators containing only $\ge 0$ powers of $\partial_2$. 

Consider the space $E^2$ and consider the Lax system (6) on the page 14 in \cite{Pa}, namely, if $N\in E^2$, then it looks like 
\begin{equation}
\label{Par}
\frac{\partial N}{\partial t_k}=V_N^k,
\end{equation}
where 
$$
V_N^k=([(L^nM^m)_{+},L], [(L^nM^m)_{+},M])
$$
if $N=(L,M)$ and $k=(n,m)$, $n,m\ge 0$. Note that we concider $N$ as belonging to the 
extended phase space $\tilde{E}^2$, where 
$\tilde{E}= A[[\ldots ,t_k,\ldots ]]((\partial_1^{-1}))((\partial_2^{-1}))$
 (to clarify $t$-dependence of operators, see section 3). Independently of $t$-dependence of operators, we can nevertheless prove some fundamental facts about the system. The first one is the following proposition. 

\begin{prop}
\label{comm}
Suppose that $\ord(LM) > 0$ and $N=(L,M)$ satisfy the Lax system. Suppose also that the highest coefficients of $L,M$ are not zero divisors. 

Then $[L,M]=0$.
\end{prop} 
 
{\bf Proof.}  Since the highest coefficients of $L,M$ are not zero divisors, we have $\ord(LM)=\ord(L)+\ord(M)> 0$, and therefore either $\ord(L)> 0$ or $\ord(M)>0$. Without loss of 
generality assume $\ord(M)>0$. Then for $k=(1,n)$ with arbitrary large $n$ we have:
$$
\frac{\partial M}{\partial t_k}=([L,M])M^n-[(LM^n)_{-},M]
$$
where from 
$$
([M,L])M^n\in E_{\ord(M)} 
$$
and $[M,L]\in E_{\ord(M)(1-n)}$.  
So, $[M,L]=0$.\\
$\Box$

Further we will study some modifications of this system and we will look for solutions $N$ 
with the property $\ord(M)=1$, $\ord(L)=0$. The propsition will remain true for all these 
systems because all these systems will depend on times indexed by infinite many different indices $k=(\ldots ,j)$ and $j$ will appear as a power of $M$. 

Now, though it will not be important for the rest of the paper, for completeness of our reseach let's consider what happens if we will try to find solutions with $\ord(M)=\ord(L)=1$. Let's consider the following example. 

\bigskip

{\bf Example.} Let us try to find  a solution of the system in the form:
$$
L=\partial +u_1\partial^{-1}+u_2\partial^{-2}+\ldots
$$
$$
M=\partial +v_1\partial^{-1}+v_2\partial^{-2}+\ldots
$$
where $\partial=\partial_2$ and $u_i,v_i\in k((x_1))((x_2))[[\ldots ,t_{ij},\ldots ]]((\partial_1^{-1}))$. 
 First of all, note that the conditions
$$
[(L^nM^m)_{+},L]_i=0, \mbox{\quad} [(L^nM^m)_{+},M]_i=0,
$$
where $i\ge 0$ and $[.]_i$ denotes the $i$-th coefficient of  an operator, are exactly the conditions for the operators $L,M$ to commute, as we have seen above. It is not difficult to prove that for $L,M$ these conditions are the following: $[L,M]=0$ iff for all $n\ge 0$ holds
\begin{equation}
\label{usloviya}
D(u_n-v_n) + \sum_{i,j\ge 1, i+j=n} [v_i, u_j] + \sum_{i,j,k\ge 1, i+j+k=n}(-1)^j C_{j+i-1}^{i-1}(v_i D^{(j)}(u_k) - u_i D^{(j)}(v_k)) = 0, 
\end{equation}
where $D()=\partial /\partial (x_2)$ and $C_i^j$ are the binary coefficients. 

The equations for $k=(1,n)$, $k=(n,1)$ for all $n$ give us the following property:
$$
\frac{\partial u_1}{t_k}=\frac{\partial v_1}{t_k}
$$ 
Indeed, this follows from (\ref{usloviya}) and the property: 
$[(L^nM^m)_{+},L]=[(L^nM^m),L]-[(L^nM^m)_{-},L]$. Therefore, $u_1-v_1=:c\in k((x_1))((\partial_1))$.

Now, the condition for the Lax system to have a solution appears if we will consider equations for  $u_l,v_l$ for $k=(i,j)$, $i+j+l\le 4$. 

Let's denote by $D$ a derivation by $x_2$, and by $u_{22}$ the derivation $D(u_2)$. We have 
$$
\frac{\partial u_1}{\partial t_{1,1}}=
\left (D^{\left (2\right )}\right )(u_{{1}})+v_{{1}}u_{{1}}+2\,
\mbox {D}(u_{{2}})-u_{{1}}v_{{1}}
$$
$$
\frac{\partial v_1}{\partial t_{1,1}}=
u_{{1}}v_{{1}}+\left (D^{\left (2\right )}\right )(v_{{1}})+2\,
\mbox {D}(v_{{2}})-v_{{1}}u_{{1}}
$$
where from 
$$
2u_{22}+[v_1,u_1]=\frac{\partial u_1}{\partial t_{11}}
$$
Then,
$$
\frac{\partial u_2}{\partial t_{1,1}}=
\left (D^{\left (2\right )}\right )(u_{{2}})+[u_{{1}},u_{{2}}]+[v_{{1}},
u_{{2}}]+2\,\mbox {D}(u_{{3}})+2u_{{1}}(\mbox {D}(u_{{1}}))
$$
$$
\frac{\partial v_2}{\partial t_{1,1}}=
2v_{{1}}(\mbox {D}(u_{{1}}))+[v_{{1}},v_{{2}}]+\left (D^{\left (2\right )}
\right )(v_{{2}})+2\,\mbox {D}(v_{{3}})+[u_{{1}},v_{{2}}]
$$
where from 
$$
\frac{\partial (u_2-v_2)}{\partial t_{1,1}}=[D(u_{1}),c]+[c,u_2]+[c,v_2]
$$
($c$ were defined above).
Differentiating by $x_2$ we get 
$$
[\frac{\partial u_1}{\partial t_{1,1}}-D^{(2)},c]=0
$$
From other equations: 
\begin{multline*}
\frac{\partial u_1}{\partial t_{1,2}}=
3\,\mbox {D}(u_{{3}})+\left (D^{\left (3\right )}\right )(u_{{1}})+2\,
[v_{{2}},u_{{1}}]+3\,\mbox {D}(v_{{1}})(u_{{1}})+\\
3\,\left (D^{\left (2
\right )}\right )(u_{{2}})+2\,v_{{1}}(\mbox {D}(u_{{1}}))+2\,[v_{{1}},u
_{{2}}]+u_{{1}}(\mbox {D}(u_{{1}}))
\end{multline*}
\begin{multline*}
\frac{\partial v_1}{\partial t_{1,2}}=
3\,\mbox {D}(v_{{3}})+\left (D^{\left (3\right )}\right )(v_{{1}})+3\,
\left (D^{\left (2\right )}\right )(v_{{2}})+\\
3\,\mbox {D}(v_{{1}})(v_{
{1}})+2v_{{1}}(\mbox {D}(v_{{1}}))+[u_{{1}},v_{{2}}]+u_{{1}}(\mbox {D}(v
_{{1}}))+[u_{{2}},v_{{1}}]
\end{multline*}
\begin{multline*}
\frac{\partial u_1}{\partial t_{2,1}}=
\left (D^{\left (3\right )}\right )(u_{{1}})+3\,\mbox {D}(u_{{3}})+2\,
u_{{1}}(\mbox {D}(u_{{1}}))+v_{{1}}(\mbox {D}(u_{{1}}))+\\
[v_{{1}},u_{{2}
}]+[v_{{2}},u_{{1}}]+3\,\mbox {D}(v_{{1}})(
u_{{1}})+3\,\left (D^{\left (2\right )}\right )(u_{{2}})
\end{multline*}
\begin{multline*}
\frac{\partial v_1}{\partial t_{2,1}}=
v_{{1}}(\mbox {D}(u_{{1}}))+3
\,\mbox {D}(v_{{3}})+\left (D^{\left (3\right )}\right )(v_{{1}})+3\,
\left (D^{\left (2\right )}\right )(v_{{2}})+\\
2\,u_{{1}}(\mbox {D}(v_{{
1}}))+2\,[u_{{1}},v_{{2}}]+3\mbox {D}(u_{{1}})(v_{{1}})+2\,[u_{{2}},v_{{1
}}]
\end{multline*}
we get 
$$
\frac{\partial u_1}{\partial t_{1,2}}-\frac{\partial u_1}{\partial t_{2,1}}=
[v_2,u_1]+[v_1,u_2]-cD(u_1)
$$
This expression includes $u_2,v_2$, and all other equations with $j+i+l>4$ will include higher terms, which means that we get no equations on $u_1$. 

This explains us that the system from this example contain not enough equations. This is, in fact, the corollary of taking the special form of solutions.   

\bigskip

\subsection{Parshin's KP-hierarchy}
A more interesting question is:  are there solutions of the system (\ref{Par}) of the form
\begin{equation}
\label{L}
L=u_0+u_1\partial_2^{-1}+\ldots
\end{equation}
\begin{equation}
\label{M}
M=v_{-1}\partial_2+v_{0}+v_1\partial_2^{-1}+\ldots
\end{equation}
where $u_0, v_{-1}$ are monic series with the orders 
$\ord_{\partial_1}(u_0)=1$, $\ord_{\partial_1}(v_{-1})=0$. 

This question was posed in one particular case in \cite{Pa}. Also, as it will be shown in section 6, it arises by studying the existence of a universal family of isospectral deformations of a pair of differential operators in two variables. Also it is related to the question posed in Remark 1.7. in \cite{Mu}. By proposition \ref{comm} we must have $[L,M]=0$, so,in particular, $[u_0,v_{-1}]=0$. 

Note that, since $\frac{\partial L}{\partial t_{ij}}=[(L^iM^j)_+,L]=-[(L^iM^j)_-,L]$, we have $\partial /
\partial t_{ij} (u_0)=0$ and analogously $\partial / \partial t_{ij} (v_{-1})=0$. So, $u_0,v_{-1}$ do not depend on times and therefore coincide with the first coefficients of the initial data. Since $u_0,v_{-1}$ are invertible operators, the operators $L,M$ of a solution will be also invertible.

Below we will prove that there are no nontrivial solutions of the form (\ref{L}), (\ref{M}) of 
our Lax system. By triviality of solutions we mean 
solutions representable either in the form $(L_0,M_0)$ with $L_0=(L_0)_{+}$, $M_0=(M_0)_{+}$ or in the form $(L,M)$, where $L,M$ are series in variables $L_0^{-1}, M_0^{-1}$ with constant coefficients. Obviously, all such series satisfy our system  of equations. Note that all such solutions don't depend on time. 

Assume the converse, i.e. that there exist non-trivial solutions. Consider the series of equations for $k=(n,0)$ and $k=(n,1)$. For 
the series $k=(n,0)$ we have 
\begin{equation}
\label{2004}
\frac{\partial L}{\partial t_k}=
\sum_{i=0}^{\infty }([u_0^n,u_i]\partial_2^{-i}+u_i[u_0^n,\partial_2^{-i}])
\end{equation}
\begin{equation}
\label{2005}
\frac{\partial M}{\partial t_k}=
\sum_{i=-1}^{\infty }([u_0^n,v_i]\partial_2^{-i}+v_i[u_0^n,\partial_2^{-i}])
\end{equation}

Let's point out the following easy observation: since $\frac{\partial }{\partial t_k}$ and $[(L^iM^j)_+,.]$ are derivations, any series in variables $L,M$ with constant coefficients satisfy all equations of our system if $(L,M)$  is a solution of the system. Now, since our solutions are assumed to be nontrivial, there exists an index $i>0$ such that after replacing 
$L$ and $M$ with $L'=L+$ (some series in $L^{-1},M^{-1}$ with constant coefficients) (correspondingly $M'=M+ \ldots$), we can assume that $u'_i, v'_i$ are the first nonzero coefficients of $L',M'$ between  all coefficients with indices $j>0$, and that $u'_i,v'_i$ are series in $\partial_1$ with {\it nonconstant} first coefficients.

From equation (\ref{2004}) we derive that $\partial u'_i/ \partial t_{n,0}= [u_0^n,u'_i]$ for all $n$. Since the order of the left hand side is bounded from above, we obtain $[u_0^n,u'_i]=0$, hence $[u_0,u'_i]=0$. Analogously $[u_0,v'_i]=0$. Using arguments from much more general lemma \ref{nichts} below we obtain  $[u'_i, v_{-1}]=0$ and $[v'_i, v_{-1}]=0$ (our situation is much more simple than in \ref{nichts}, because we deal with the ring $\tilde E$ of usual, though time-dependent, pseudo-differential operators). 
Also we get that $u_0, v_{-1}, u'_i,v'_i$ do not depend on $t_{n,0}$ for all $n$ and that the first coefficients of $u'_i,v'_i$ belong to $\ker\partial_1$. Using the same arguments we obtain from (\ref{2004}), (\ref{2005}) that 
\begin{equation}
\label{tratata}
[u_0, u'_{i+1}]+iu'_i\partial_2(u_0)=0, \mbox{\quad}
[u_0, v'_{i+1}]+iv'_i\partial_2(u_0)=0.
\end{equation}

Now consider the series of equations for $k=(n,1)$. 

\begin{lemma}
\label{LM_+}
We have 
$$
(L^nM)_+=u_0^nv_{-1}\partial_2+nu_1u_0^{n-1}v_{-1}+u_0^nv_{0}
$$
\end{lemma} 

{\bf Proof.} By induction on $n$. For $n=0$ we have $M_{+}=v_{-1}\partial_2+v_0$. 
Since $L$ contains only nonpositive order terms, and $L^{n-1}M$ has only one positive order term, namely $u_0^{n-1}v_{-1}\partial_2$, the plus-part of the operator $L^nM$ will be 
\begin{equation}
\label{recursion}
(L^nM)_+=u_1u_0^{n-1}v_{-1} +u_0((L^{n-1}M)_+)
=
u_0^nv_{-1}\partial_2+nu_1u_0^{n-1}v_{-1}+u_0^nv_{0}
\end{equation}
$\Box$

Now for $k=(n,1)$  we have by formula (\ref{recursion})
$$
\frac{\partial u'_i}{\partial t_{n,1}}=
[(L^nM)_+,L]_i=[u_0^nv_{-1}\partial_2+nu_1u_0^{n-1}v_{-1}+u_0^nv_{0}, L]_i=
$$
$$
[u_0^nv_{-1}\partial_2, u'_i\partial_2^{-i}]_i + [u_0^nv_{-1}\partial_2, u'_{i+1}\partial_2^{-i-1}]_i
+ [nu_1u_0^{n-1}v_{-1}+u_0^nv_{0}, u'_i\partial_2^{-i}]_i=
$$
$$
([u_0^nv_{-1}, u'_i\partial_2^{-i}]\partial_2)_i+ u_0^nv_{-1}\partial_2(u'_i)+ [u_0^nv_{-1}, u'_{i+1}]
+u_0^n[v_0,u'_i]=
$$
$$
iu'_i\partial_2(u_0^nv_{-1})+ u_0^nv_{-1}\partial_2(u'_i)+ u_0^n[v_0,u'_i] +u_0^n[v_{-1}, u'_{i+1}]- iu'_i\partial_2(u_0^n)v_{-1}= 
$$
\begin{equation}
\label{nif}
u_0^n(iu'_i\partial_2(v_{-1})+ v_{-1}\partial_2(u'_i) + [v_0,u'_i] + [v_{-1}, u'_{i+1}]), 
\end{equation}
where the transformations follows from the fact that $[u_1,u'_i]=0$ ($u_1$ is either equal to $u'_i$ or is representable as a series in $u_0$ with coefficients belonging to the ring 
$k[[\ldots ,t_{ij}, \ldots ]]$ multiplied by $v_{-1}^{-1}$ by the induction hypothesis). So, by usual arguments  we obtain ${\partial u'_i}/{\partial t_{n,1}}=0$, 
\begin{equation}
\label{nuf}
iu'_i\partial_2(v_{-1})+ v_{-1}\partial_2(u'_i) + [v_0,u'_i] + [v_{-1}, u'_{i+1}]=0
\end{equation}
and analogously ${\partial v'_i}/{\partial t_{n,1}}=0$,
\begin{equation}
\label{naf}
iv'_i\partial_2(v_{-1})+ v_{-1}\partial_2(v'_i) + [v_0,v'_i] + [v_{-1}, v'_{i+1}]=0.
\end{equation}
Assume that $\ord_{\partial_1}(\partial_2(u'_i))=\ord_{\partial_1}(u'_i)$, that is the first coefficient of $u'_i$ does not belong to the ring $k[[\ldots ,t_{ij}, \ldots ]]$. Then 
$\ord_{\partial_1}(v_{-1}\partial_2(u'_i))=\ord_{\partial_1}(u'_i)$. 

Let's compare the orders of other summands in formula (\ref{nuf}). 
Since $u_0$ is monic, we can write $u'_i$ as a series in $u_0^{-1}$. These series will have coefficients belonging to $\ker\partial_1$, because $[u'_i,u_0]=0$. The same is true also for the operator $v_{-1}u_0$. This observation imply that the orders of commutators with these operators will be less or equal to 
the orders of commutators with $u_0^{\ord_{\partial_1}(x)}$, where $x=u'_i$ or $x=v_{-1}u_0$. 
From equation (\ref{2005}) we obtain (as in (\ref{tratata})) that $[u_0,v_0]-v_{-1}\partial_2(u_0)=0$, where from 
$$
\ord_{\partial_1}([v_0,u'_i])\le  \ord_{\partial_1}([v_0,u_0^{\ord_{\partial_1}(u'_i)}])=\ord_{\partial_1}(v_{-1}\partial_2(u_0^{\ord_{\partial_1}(u'_i)})) < \ord_{\partial_1}(u'_i).
$$
Now we have 
$$\ord_{\partial_1}(iu'_i\partial_2(v_{-1}))< \ord_{\partial_1}(u'_i)$$ 
because of moniqueness of $u_0$, 
$$
\ord_{\partial_1}([v_{-1}, u'_{i+1}])=\ord_{\partial_1}([v_{-1}u_0, u'_{i+1}]u_0^{-1}+ v_{-1}u_0[u_0^{-1},u'_{i+1}])\le  
$$
$$
\max\{\ord_{\partial_1}([u_0,u'_{i+1}]u_0^{-1}),
\ord_{\partial_1}(v_{-1}[u_0,u'_{i+1}]u_0^{-1})\}=
$$
$$
\max\{\ord_{\partial_1}(u'_i\partial_2(u_0)u_0^{-1}), 
\ord_{\partial_1}(v_{-1}u'_i\partial_2(u_0)u_0^{-1})\}< \ord_{\partial_1}(u'_i)
$$

So, our assumption contradicts with equation (\ref{nuf}). Denote the first coefficient of $u'_i$ by $c$ and consider the operator $L''= L' - c{L}^{\ord_{\partial_1}(u'_i)}{M}^{-i}$. As we have proved, $c\in k[[\ldots ,t_{ij},\ldots ]]$.
Since $c$ does not depend on $t_{n,0}, t_{n,1}$, the operator $L''$ satisfy the equations (\ref{2004}) and (\ref{nif}) and $u''_i=u'_i-cu_0^{\ord_{\partial_1}(u'_i)}v_{-1}^{-i}$ has the order $\ord_{\partial_1}(u''_i)<\ord_{\partial_1}(u'_i)$. Repeating all the above arguments we conclude that $u'_i$ can be written as  a series in $u_0$ with coefficients in $k[[\ldots ,t_{ij}, \ldots ]]$ multiplied by $v_{-1}^{-i}$. 
Clearly, the same conclusion is true for $v'_i$. 

Continuing this line of reasons we get that ${L'}_+=L_+$ and ${M'}_+=M_+$ can be written as series in $L,M$ with coefficients in $k[[\ldots ,t_{ij}, \ldots ]]$ that don't depend on $t_{n,0}, t_{n,1}$, where from $L,M$ can be written as series in $L_+,M_+$ with such coefficients. Since $[u_0,v_{-1}]=0$, $[u_0,v_0]-v_{-1}\partial_2(u_0)=0$, the operators $L_+,M_+$ commute. Therefore, $[(L^iM^j)_+,N]=0$ for all $i,j$, where from we get $\partial N/ \partial t_{ij}=0$ for all $i,j$. So, $(L,M)$ must be a trivial solution. Combining all together, we obtain 

\begin{prop}
\label{trivialitaet}
The system (\ref{Par}) has no nontrivial solutions of the form (\ref{L}), (\ref{M}) in any ring $\tilde{E}$. 
\end{prop}

The proposition can be even more generalized, see remark after corollary \ref{corollary}, section 3.2.


\section{Modified Parshin's KP-hierarchy}

\subsection{General setting}

Now we introduce the following {\bf{\it modified}} Lax systems:
$$
\frac{\partial N}{\partial t_k}=V_N^k,\eqno{(KP)_{\alpha}}
$$
where 
$$
V_N^k=([(L^nM^m)_{+},L], [(L^nM^m)_{+},M])
$$
if $N=(L,M)$ and 
$k=(i,j)$, $j\ge 0$, $i\le \alpha j$, $i\in\dz$, and $\alpha$ is any function $\alpha :\dz_+\rightarrow \dr$ such that $\alpha (0)\le 0$. Here and below we write $\alpha j$ instead of $\alpha (j)$. The main example of such function is a linear function, $j\mapsto \alpha j$ with $\alpha\in\dr$.  

We will look for a solution $(L,M)$ of these systems with $\ord_{\partial_2}(L)=0$, $\ord_{\partial_2}(M)=1$ and with initial conditions 
$$
L_0=u_0+u_1\partial_2^{-1}+\ldots
$$
$$
M_0=v_{-1}\partial_2+v_0+v_1\partial_2^{-1}+\ldots ,
$$
where $u_i,v_i\in A((\partial_1^{-1}))$ are monic operators with  
$\ord_{\partial_1}(u_0)=1$, 
$\ord_{\partial_1}(v_{-1})=0$.
As it will be explained in section 6, we can consider even more narrow set of initial conditions by assuming that 
$v_{-1}= \partial_1u_0^{-1}$, 
$(v_0)_{-}\cap ker( \partial_1)=0$, where the last condition means that all monomials in all coefficients of $(v_0)_{-}$ do not belong to $ker(\partial_1)$.

From now on and until the end of the article we additionally assume that the following short sequences are exact: 
$$
A\arrone A\rightarrow 0, \mbox{\quad} A\arrtwo A\rightarrow 0, \mbox{\quad} \ker \partial_2\arrone \ker \partial_2\rightarrow 0. 
$$
It is easy to see that these conditions imply also the exactness of the sequence $\ker \partial_1 \arrtwo \ker \partial_1 \rightarrow 0$.

To clarify the notion of the ring to which the coefficients of solutions belong let's introduce 
the following notation. 

First consider the ring $A_t:= A[\ldots ,t_{ij},\ldots ]$ of polynomials in infinite number of variables $t_{ij}$, $i\in\dz$, $j\in\dz_+$ with coefficients from the ring $A$. We assume that variables commute with each other and with elements of $A$. Let's define a pseudo-valuation $v$ on this ring,
$$
v: A_t\backslash \{0\}\longrightarrow \dz\oplus \dz_+
$$ 
by $v(t_{i,j})=(-i,j)$, $v(a)=0$ for $a\in A$. We define also a pseudo-valuation $v_2: A_t\backslash \{0\}\longrightarrow \dz_+$ by $v_2(t_{ij})=j$. We assume here $(i_1,j_1)>(i,j)$ if $j_1>j$ or $j_1=j$ and $i_1>i$. 
Recall that a pseudo-valuation $\nu$ on the ring $R$ with values in an ordered abelian group $\Gamma$ is a function 
$$
\nu :R\rightarrow \Gamma\cup\{\infty\}
$$
such that $\nu (0)=\infty$, $\nu (ab)\ge \nu (a)+\nu (b)$ and $\nu (a+b)\ge \min\{\nu (a),\nu (b)\}$. To simplify the terminology further in this paper we will identify the words {\it valuation} and {\it pseudo-valuation}. 

Now we introduce a group topology on $A_t$ considering $A_t$ as an abelian group. This topology is an appropriate model of  the topology on a two-dimensional local field (see \cite{PaFi} for all details concerning the topologies on higher local fields, or also \cite{Ye}, \cite{Pa2}).
Namely, we define the base of neigbourhoods of zero as the set of all sets of the following type 
$$
U:=\{\sum u_i\in A_t \mbox{ with $v_2(u_i)=i$ and } v(u_i)>(j_i,i)\},
$$
where $\{j_i\}$ is a system of integer numbers with $j_i=-\infty$ for large $i$. 

The completion $\bar{R}:=\hat{A_t}$ of the topological group $A_t$ with respect to this topology has a  structure of an associative  $k$-algebra with the componentwise multiplication of fundamental sequences. Every element of this algebra can be thought of as a series, whose summands are monomials belonging to $A_t$, such that every neigbourhood of zero in $A_t$ contains almost all summands of the series. The valuation $v$ (and $v_2$) can be uniquely extended to the ring $\bar{R}$ by the rule 
$$
v(\sum a_i)=\min\{v(a_i)\},
$$
where $\{a_i\}\in A_t$ are monomials.  We extend the derivations $\partial_1,\partial_2$ to the ring $\bar{R}$ in a usual way by assuming that all $t_{ij}\in \ker\partial_1\cap\ker\partial_2$. Now define 
$$
E_{\bar{R}}:=\bar{R}((\partial_1^{-1})),\mbox{\quad}
E_{R}:=E_{\bar{R}}((\partial_2^{-1})),\mbox{\quad}
 E_k:=k((\partial_1^{-1}))((\partial_2^{-1})), 
$$
$$
\Nu_{\bar R}:=\{1+{E_{\bar R}}_{-}\},
\mbox{\quad}
\Nu_R:=\{1+{E_R}_{-}\},
$$
where the decomposition for the ring $E_R$ in plus and minus parts is defined in the same way as in section 2.1, and the decomposition for the ring $E_{\bar R}$ is defined analogously with respect to $\partial_1$.

We extend the valuation $v_2$ from $\bar{R}$ to $E_{\bar{R}}$ by $v_2(\sum a_k\partial_1^k)=\min\{v_2(a_k)\}$. One can check immediately that this definition is correct. Now we can give an appropriate definition of the ring $\bar{R}\{\{\partial_1^{-1}\}\}$: we define 
\begin{multline*}
\widehat{E}_{\bar{R}}:=\{L=\sum_{q\in\sdz}b_q\partial_1^q | b_q\in\bar{R}\mbox{ {\it and 
for any integer} $M$ {\it and positive integer} $N$}\\
\mbox{ {\it there exist only finite number of} $b_q$ {\it with} $q<M$ {\it such that} $v_2(b_q)=N$}\}
\end{multline*}

\begin{lemma}
\label{kol'co1}
The set $\widehat{E}_{\bar{R}}$ is a ring. 
\end{lemma}

{\bf Proof.} Obviously, the set $\widehat{E}_{\bar{R}}$ is an abelian group. The multiplication of two series is defined by the same formula as for the ring $E_{\bar{R}}$. We must check only that it is well defined and the product of two series belong again to $\widehat{E}_{\bar{R}}$. 

For two series $A=\sum_{q\in\sdz}a_q\partial_1^q$, $B=\sum_{q\in\sdz}b_q\partial_1^q$ we have 
$$
AB=\sum_{q\in\sdz}g_q\partial_1^q,
$$
where
\begin{equation}
\label{star}
g_q=\sum_{k\in\sdz}\sum_{l\ge 0}C_k^la_k\partial_1^l(b_{q+l-k}).
\end{equation}
By definition of the set $\widehat{E}_{\bar{R}}$ for any integer $M$ and positive integer $N$ there exist integer $M_1,M_2$ such that $v_2(a_k)>N$ for any $k>M_1$ and $v_2(b_q)>N$ for any $q>M_2$. Since $v_2(\partial_1(b_q))\ge v_2(b_q)$, we obtain that all summands in (\ref{star}) for any $k> M_1$ and arbitrary $l$ or for $k\le M_1$ and $l>M_2+k-q$ have valuation greater than $N$. So, the number of summands with valuation {\it less} than $N$ is finite. Therefore, the series in (\ref{star}) converges for any $q$. 
Moreover, if we take $q>M_2+M_1$ then we obtain $v_2(g_q)>N$, where from we get that $AB\in \widehat{E}_{\bar{R}}$. 

The associativity and distributivity can be easily deduced in the same way as for the ring  ${E}_{\bar{R}}$.\\
$\Box$

We extend the valuation $v_2$ from ${E}_{\bar{R}}$ to $\widehat{E}_{\bar{R}}$ in the same way as for the ring ${E}_{\bar{R}}$. One can check immediately that this definition is also correct.

Now we give an appropriate definition of the ring $\bar{R}\{\{\partial_1^{-1}\}\}\{\{\partial_2^{-1}\}\}$: we define 
\begin{multline*}
\widehat{E}_R:=\{L=\sum_{q\in\sdz}a_q\partial_2^q | a_q\in\widehat{E}_{\bar{R}}\mbox{ {\it and there is a positive real number} $C_L$ {\it and positive} }\\ 
\mbox{{\it integer} $M_L$ {\it such that} $v_2(a_q)>C_Lq$ {\it for all} $q>M_L$}\} ,
\end{multline*}

\begin{lemma}
\label{kol'co2}
The set $\widehat{E}_{{R}}$ is a ring. 
\end{lemma}

{\bf Proof.} The proof is analogous to the proof of lemma \ref{kol'co1}. 
Obviously, the set $\widehat{E}_{{R}}$ is an abelian group. The multiplication of two series is defined by the same formula as for the ring $E_{{R}}$. We must check only that it is well defined and the product of two series belong again to $\widehat{E}_{{R}}$. 

For two series $A=\sum_{q\in\sdz}a_q\partial_2^q$, $B=\sum_{q\in\sdz}b_q\partial_2^q$ we have 
$$
AB=\sum_{q\in\sdz}g_q\partial_2^q,
$$
where
\begin{equation}
\label{star1}
g_q=\sum_{k\in\sdz}\sum_{l\ge 0}C_k^la_k\partial_2^l(b_{q+l-k}).
\end{equation}

First we have to check that $g_q$ belong to the ring $\widehat{E}_{\bar{R}}$. Since the linear growth condition on the valuation of coefficients is stronger than the growth condition from the definition of the ring $\widehat{E}_{\bar{R}}$, we obtain, as in lemma \ref{kol'co1}, that for any positive integer $N$ there are only finite number of summands in (\ref{star1}) with valuation less than $N$. This means that this condition holds for each $i$th coefficient of each summand. So, $g_q=\sum_ig_{qi}\partial_1^i$ with $g_{qi}\in \bar{R}$. Since there are only finitely many summands in (\ref{star1}) with valuation less than $N$, for any integer $M$ there exist only finite number of $g_{qi}$ with $i<M$ and $v_2(g_{qi})\le N$. Therefore, 
$g_q\in \widehat{E}_{\bar{R}}$ for any $q$.  

Second we have to check that $v_2(g_q)$ satisfy some linear growth condition. But this is true because of general construction of the ring as it was made in \cite{Mu}. Namely, if $M_A,C_A,M_B,C_B$ are the numbers from the definition of the set  $\widehat{E}_{{R}}$ for the elements $A,B$, then $M_{AB}=M_A+M_B$ and $C_{AB}=(\min\{C_A,C_B\})(\min\{M_A,M_B\})/(M_A+M_B)$ can be taken as the corresponding numbers for the element $AB$. 

The rest of the proof is clear.\\
$\Box$

The valuation $v_2$ can be obviously extended also to the ring $\widehat{E}_{{R}}$. Now let's define the following rings and groups (analogs of corresponding objects from \cite{Mu} defined in one dimensional case):
$$
D_R:=\{ L=\sum_{n=0}^{\infty} a_n\partial_2^n \mbox{\quad }|\mbox{\quad } a_n\in \widehat{E}_{\bar R},\mbox{\quad } a_n=0 \mbox{\quad for $n\gg 0$}\} ,
$$
$$
\widehat{D}_R:=\{L\in \widehat{E}_R | L_{-}=0\}, \mbox{\quad} \widehat{\Nu}_R:=\{1+{\widehat{E}_{R-}}\},
$$ 
$$
\widehat{\Nu}_{k}:=\{L=1+\sum_{i=1}^{\infty}a_i\partial_2^{-i}| a_i=\sum_{q\in\sdz}b_q\partial_1^q\in \widehat{E}_{\bar{R}} \mbox{\quad {\it and} $b_q\in k[[\ldots ,t_{ij},\ldots ]]$}\},
$$  
where $k[[\ldots ,t_{ij},\ldots ]]$ is a subring of $\bar R$ defined in the same way;
$$
{\widehat{E}_R}^{\times}:=\{L\in \widehat{E}_R| L|_{t=0}\in \Nu_{E} \mbox{\quad and } (\pi (L))_+\in \Nu_{\bar R} \},  
$$
$$
{\widehat{D}_R}^{\times}:=\{L\in \widehat{D}_R| L|_{t=0}=1 \mbox{\quad and } (\pi (L))_+\in \Nu_{\bar R} \},
$$
where $\pi :\widehat{E}_{{R}}\rightarrow \widehat{E}_R/I$ is a projection, $I$ is an ideal of valuation $v_2$ in $\widehat{E}_R$.  

\begin{lemma}
\label{3.1}
Every element $P\in {\widehat{E}_R}^{\times}$ is invertible in ${\widehat{E}_R}^{\times}$. More precisely, the Neumann series $\sum_{n=0}^{\infty}(1-P)^n$ gives a well-defined element in ${\widehat{E}_R}^{\times}$, which we denote $P^{-1}$. 
\end{lemma}

The proof of lemma will be given later. Now let's note that by this lemma ${\widehat{E}_R}^{\times}, {\widehat{D}_R}^{\times}$ are groups. Indeed, using formulas  (\ref{star1}), (\ref{star}) one can easily show that the product of two elements belong again to the same set. All other group laws follow from the  laws of the ring $\widehat{E}_R$.

Using arguments of proposition \ref{comm}, one gets that there can exist a nontrivial solution only if $[L_0,M_0]=0$. So, {\it we will assume until the end of section that $L_0,M_0$ commute}. It is easy to see that the modified systems satisfy all the basic properties derived in \cite{Pa} for the original system.

\subsection{Technical tools}

In this subsection we prove several important technical facts, which will be used later. These facts generalize well-known results from \cite{Pa}, \cite{Mu}. 

The special choice of the coefficients $u_0,v_0, v_{-1}$ mentioned in the beginning of section 3.1 is partially explained by the theorem below, which is a generalization of theorem 1 in \cite{Pa}. Another part of the explanation will be given in section 6.  

\begin{theo}
\label{Parshin}
Let $L,M\in {\widehat E}_R$  be two monic operators satisfying the following condition:\\
$\ord_{\partial_2}(L)=0$, $\ord_{\partial_1}(L_+)=1$, $\ord_{\partial_2}(M)=1$, 
$\ord_{\partial_1}(({M\partial_2^{-1}})_+)=0$. \\
Then 

(i) $[L,M]=0$ if and only if there exists an  operator $S\in {\widehat \Nu}_R$ such that 
$$
L=S^{-1}L_+S, \mbox{\quad} M=S^{-1}(({M\partial_2^{-1}})_+\partial_2 +v_0)S,
$$
where $v_0$ is a uniquely defined by $L,M$ element and $[L_+,({M\partial_2^{-1}})_+\partial_2+v_0]=0$. 

ii) If $S,S'$ are two operators from i) then 
$$
S'S^{-1}\in 1+k[[\ldots ,t_{ij},\ldots ]]((L_{00}^{-1}))((M_{00}^{-1}))\cap {\widehat E}_{R-} ,
$$
where $L_{00}=L_+$, $M_{00}=({M\partial_2^{-1}})_+\partial_2 +v_0$.

iii) Suppose that the equation $\partial_1(x)=cx$  is solvable for any $c\in A$. 
Then 
there exists a zeroth order invertible operator $\bar{S}\in {\widehat E}_{\bar R}$ such that
$$
L_+={\bar{S}}^{-1}u_0\bar{S}, \mbox{\quad} (({M\partial_2^{-1}})_+\partial_2 +v_0)={\bar{S}}^{-1}(v_{-1}\partial_2+\tilde{v_0})\bar{S},
$$
where  $u_0,v_{-1}$ are monic, $\ord_{\partial_1}(u_0)=1$, 
$\ord_{\partial_1}(v_{-1})=0$,
$v_{-1}=\partial_1u_0^{-1}$, 
$(\tilde{v_0})_{-}\cap ker( \partial_1)=0$.

If $\bar{S'}$ is another operator with these properties, then $\bar{S}\bar{S'}^{-1}$ is an operator with  coefficients from $k[[\ldots ,t_{ij},\ldots ]]$. 
\end{theo}

{\bf Proof.} i) The "if" part is clear. To prove the "only if" part we will need the following lemmas:

\begin{lemma}
\label{commut}
Let $L,M\in {\widehat E}_R$ be two arbitrary operators of finite order and define $N=\max\{\ord_{\partial_2}(L), \ord_{\partial_2}(M)\}$. Then $[L,M]=0$ if and only if 
$$
\sum_{\indeks}C_i^j(v_i\partial_2^{(j)}(u_k)-v_i\partial_2^{(j)}(u_k))+\sum_{\indeksa}[v_l, u_n]=0
$$
for all $\alpha \le \ord_{\partial_2}(L)+\ord_{\partial_2}(M)$, 
where 
$L=\sum u_i\partial_2^i$, $M=\sum v_j\partial_2^j$ and $u_i,v_j$ are assumed to be zero if $i>\ord_{\partial_2}(L)$, correspondingly, $j>\ord_{\partial_2}(M)$ .
\end{lemma}

{\bf Proof.} The proof follows from easy calculations. From definition we have 
$$
ML=\sum_{i=\ord_{\partial_2}(M)}^{-\infty}\sum_{j=1}^{\infty}\sum_{k=\ord_{\partial_2}(L)}^{-\infty}
v_i\partial_2^{(j)}(u_k)C_i^j\partial_2^{i-j+k}+\sum_{l=\ord_{\partial_2}(M)}^{-\infty}\sum_{n=\ord_{\partial_2}(L)}^{-\infty}v_lu_n\partial_2^{l+n},
$$ 
$$
LM=\sum_{i=\ord_{\partial_2}(L)}^{-\infty}\sum_{j=1}^{\infty}\sum_{k=\ord_{\partial_2}(M)}^{-\infty}
u_i\partial_2^{(j)}(v_k)C_i^j\partial_2^{i-j+k}+\sum_{l=\ord_{\partial_2}(L)}^{-\infty}\sum_{n=\ord_{\partial_2}(M)}^{-\infty}u_lv_n\partial_2^{l+n},
$$
where from 
$$
[M,L]=\sum_{i=N}^{-\infty}\sum_{j=1}^{\infty}\sum_{k=N}^{-\infty}C_i^j(v_i\partial_2^{(j)}(u_k)-u_i\partial_2^{(j)}(v_k))\partial_2^{i-j+k}+\sum_{l=N}^{-\infty}\sum_{n=N}^{-\infty}[v_l,u_n]\partial_2^{l+n}
$$
and the rest of the proof is clear.\\
$\Box$

\begin{lemma}
\label{nichts}
Let $u,v,r\in {\widehat E}_{\bar R}$ be three operators such that $\ord_{\partial_1}(u)=1$
and $u$ is monic. Then the conditions $[u,v]=0$ and $[u,r]=0$ imply $[v,r]=0$. Moreover, $v$ and $r$ can be represented as series in $u^{-1}$ with coefficients from $\ker\partial_1$. 
\end{lemma}

{\bf Proof.} Clearly the first assertion follows from the second one. So, let's prove that, for example, $r$ can be  represented as a series in $u^{-1}$ with coefficients from $\ker\partial_1$. 

Since $u$ is monic, each power of $\partial_1$ can be represented as a series in $u^{-1}$, where from $r$ can be also represented as a series in $u^{-1}$, say $r=\sum b_ku^k$.  
Let $N=v_2(r)$ and let $k$ be the maximal number such that $v_2(b_k)=N$. We have 
\begin{equation}
\label{dsl}
0=[u, r]=[\partial_1,r]+[u-\partial_1, r]
\end{equation}
and if $g_k$ is a corresponding coefficient of $[u-\partial_1, r]$, then $v_2(g_k)>N$, because $\ord_{\partial_1}(u-\partial_1)\ge 0$ and the ring $A$ is commutative. From the other hand side, if we denote by $g'_k$ the corresponding coefficient of $[\partial_1,r]=\partial_1(r)$, we obtain $g'_k=\partial_1(b_k)+$ "coefficients of valuation $>N$". 

Therefore, the  equation (\ref{dsl}) implies that $v_2(\partial_1(b_k))>N$, where from we obtain that $b_k=b_{k1}+b_{k2}$, where $v_2(b_{k1})=N$, $v_2(b_{k2})>N$ and  $b_{k1}\in \ker\partial_1$. 
So, we can write $r=b_{k1}u^k+r'$, where the $k$th coefficient of $r'$ has valuation greater that $N$.

Continuing this line of reasoning we obtain that $r=r_1+r''$, where $r_1=\sum_{i=k}^{-\infty}b_{i1}u^i$ and $v_2(r'')>N$. Repeating this procedure for $r''$ and so on, we get the representation 
$r=\sum_{i=1}^{\infty} r_i$, where $v_2(r_i)=N_i$, $N_{i+1}>N_i$ and $r_i=\sum_{q=k_i}^{-\infty}s_{q}u^q$ with $s_q\in \ker\partial_1$ for all $q$. Clearly, the series $\sum_{i=1}^{\infty} r_i$ with such properties converges in ${\widehat E}_{\bar R}$, so our representation is well defined. But this is exactly what we need. \\
$\Box$


Now let 
$$
L=u_0+ L_m=u_0+u_m\partial_2^m+(<m),
$$
\begin{equation}
\label{dva}
M=v_{-1}\partial_2+v'_0 +M_{m} =v_{-1}\partial_2+v'_0+v_{m+1}\partial_2^{m+1}+(<m+1),
\end{equation}
where we assume that $v_{m+1}=0$ if $m=-1$. 
If $S=1-P$, $P=b\partial_2^m$, $b\in {\widehat E}_{\bar R}$, then it is easy to check that 
\begin{equation}
\label{formulka1}
[u_0,P]=[u_0,b]\partial_2^m-mb\partial_2(u_0)\partial_2^{m-1}+(<m-1), 
\end{equation}
\begin{equation}
\label{formulka2}
[v_{-1}\partial_2+v'_0, P]=[v_{-1},b]\partial_2^{m+1}+(v_{-1}\partial_2(b)+[v'_0,b]+mb\partial_2(v_{-1}))\partial_2^m+(<m)
\end{equation}
Now we are ready to prove the theorem using subsequent approximations in powers of $\partial_2$. We will do this in several steps. 

{\bf a)} Let's introduce a function $F: \dz\times \dz_+ \times {\widehat E}_{\bar R} \longrightarrow \dz_+$ by the rule
$$
F(M,N,a)=q-M
\mbox{, {\it where} $q>M$ {\it is the maximal number such that the coefficient}  }
$$
$$
\mbox{$b_q$ {\it of} $a$, {\it where} $a=\sum b_ku_0^k$, {\it satisfy the property} $v_2(b_q)\le N$;}
$$
$$
 \mbox{{\it if there are no such coefficients, we put} $F(M,N,a)=0$}.
$$
The function $F$ satisfy the following properties:

i) $F(M,N,a)\ge F(M,N,\partial_1(a))$;

ii) $F(M,N,\sum_{i=1}^ka_i)\le \max\{F(M,N,a_i)\}$;

iii) Let $[x,u_0]=0$. Then for any $M, N$  
\begin{equation}
\label{FFF}
F(M,N,\partial_1(x))\le F(M,N,x),
\end{equation}
where the equality holds only if $F(M,N,x)=0$.

The first two properties are obvious. Let's prove the third one. For a given $M, N$ let $q$ be the number mentioned in the definition of $F$. 
By  lemma \ref{nichts} $b_q\in \ker\partial_1$, so we obtain 
$v_2(\partial_1(b_k))>N$. Hence all the coefficients $b'_k$ with $k\ge q$ of $\partial_1(x)$, where $\partial_1(x)$ is represented as a series in $u_0$, have valuations greater than $N$,  where from we obtain (\ref{FFF}).

{\bf b)} Let the operators $L,M$ have the form as in (\ref{dva}). We will look for an $S$ of the form as above. We have
$$
S^{-1}LS=(1+P+P^2+\ldots )L(1-P)=L-[L,P]-P[L,P]-P^2[L,P]-\ldots =
$$
$$
L-[L,P]-P[L,P]-P^2[L,P]-\ldots =
$$
\begin{equation}
\label{formulka}
u_0+L_m-[u_0,P]-[L_m,P]-P[u_0,P]-\ldots 
\end{equation}
All the terms, except the first three, are the elements of the order less or equal to ${2m}$. Hence, we get 
$$
S^{-1}LS=u_0+(u_m-[u_0,b])\partial_2^m +(<m).
$$ 

To solve the equation $u_m-[u_0,b]=0$ let's define a sequence of elements in ${{\widehat E}_{\bar R}}$ as follows. Let $b_0$ be a solution of the equation $u_m=\partial_1(b_0)$. It exists because of conditions on the ring $A$, and we can always choose the coefficients of $b_0$ in such a way that values of the valuation $v_2$ on corresponding coefficients of $u_m, b_0$ are equal, so 
the conditions from the definition of the ring $ {{\widehat E}_{\bar R}}$ hold also for $b_0$. 

Let $b_1$ be a solution of the equation $-[u_0-\partial_1, b_0]=\partial_1(b_1)$, and, generally, let $b_k$ be a solution of the equation $-[u_0-\partial_1, b_{k-1}]=\partial_1(b_k)$. Since $\ord_{\partial_1}(u_0-\partial_1)\le 0$ and the ring $A$ is commutative, we have the following property for the elements $[u_0-\partial_1, b_{k-1}]$ and therefore for $b_k$: 
$$
F(M,N,b_k)=F(M,N,[u_0-\partial_1, b_{k-1}])\le F(M,N,b_{k-1}),
$$
where the equality holds only if $F(M,N,b_{k-1})=0$.

Therefore, the series $b:=\sum_{k=0}^{\infty}b_k$  converges in ${{\widehat E}_{\bar R}}$ and gives a solution of the equation $u_m-[u_0,b]=0$. 

{\bf c)} For the operator $M$ we have 
\begin{equation}
\label{formulka0}
S^{-1}MS= v_{-1}\partial_2+v'_0+M_m-[v_{-1}\partial_2+v'_0,P]-[M_m,P]-P[v_{-1}\partial_2+v'_0,P]-\ldots ,
\end{equation}
where from 
\begin{multline*}
S^{-1}MS=v_{-1}\partial_2+v'_0+ (v_{m+1}-[v_{-1},b])\partial_2^{m+1}+\\
(v_m- (v_{-1}\partial_2(b)+[v'_0,b]+mb\partial_2(v_{-1})))\partial_2^m +(<m).
\end{multline*}
From lemma \ref{nichts} follows that $[v_{-1},b]$ does not depend on the choice of $b$ above. Therefore, we can define $v_0:=v'_0- [v_{-1},b_{u_1}]$, where $b_{u_1}$ is a solution of the equation 
$u_1-[u_0,b]=0$. The relation for $\alpha =0$ from lemma \ref{commut} for operators $L,M$ looks like 
\begin{equation}
\label{tozhd}
0=v_{-1}\partial_2(u_0)+[v'_0,u_0]+[v_{-1},u_1]=v_{-1}\partial_2(u_0)+[v_0,u_0]
\end{equation}
and the last equation is exactly the relation for $\alpha =0$ for operators $L_+, (({M\partial_2^{-1}})_+\partial_2 +v_0)$. All relations for $\alpha <0$ are identically zero and the relation for $\alpha =1$ coincide with the relation for operators $L,M$, so the operators $L_+$ and  $(({M\partial_2^{-1}})_+\partial_2 +v_0)$ commute by lemma  \ref{commut}. 

{\bf d)} Now we can assume $m<-1$, so that $v'_0=v_0$, $u_{-1}=0$ in (\ref{dva}). Since the operators $M, L$ commute, let's consider the relation from lemma \ref{commut} for these operators for $\alpha =m+1$. We obtain
$$
[v_{m+1},u_0]= 0. 
$$
We claim that there exists an element $s\in {{\widehat E}_{\bar R}}$ such that $[u_0,s]=0$ and 
$$
v_{m+1}-(v_{-1}\partial_2(s)+[v_0,s]+(m+1)s\partial_2(v_{-1}))=[v_{-1},d],
$$
where $d$ is a solution of the equation $[u_0,d]=(m+1)s\partial_2(u_0)$. 

First of all let us note that $[v_{-1},d]$ does not depend on the choice of $d$ because of lemma \ref{nichts}, so the above equation is uniquely determined. As it follows from formulas (\ref{formulka1}), (\ref{formulka2}), (\ref{formulka}), (\ref{formulka0}),  for the operators 
$S_1:=1+s\partial_2^{m+1}$, $S_2:=1+d\partial_2^{m}$  we obtain 
$$
S_2^{-1}S_1^{-1}LS_1S_2=u_0+(<m), \mbox{\quad} S_2^{-1}S_1^{-1}MS_1S_2=v_{-1}\partial_2+v_0+(< m+1).
$$
The operators $S_i$ obtained in this way step by step can be multiplied inside the group ${\widehat \Nu}_R$ and the result will be a solution of our problem.

{\bf e)} To find $s$ we will construct a sequence of elements $s_i$ such that $[s_i,u_0]=0$ for each $i$ and the sum $s:=\sum s_i$ converges. 
 
Let $v_{m+1}=\sum b_ku_0^k$. 
Let's first define the element $s_{0}=\sum s_{0k}u_0^k$ to be a solution of the equation 
$$
\sum b_ku_0^k=\sum \partial_2(s_{0k})u_0^k.
$$
We can always find a solution of such equation because of our assumptions on the ring $A$. But we will need a special kind of solutions, namely, we will look for a solution $s_{0}$ in such a way that 

1) $v_2(b_k)=v_2(s_{0k})$ for all $k$;

2) $s_{0k}\in \ker\partial_1$ for all $k$. 

Obviously, we can find a solution satisfying both conditions by the same reason. Note that such a special choice of a solution imply the 
property (\ref{FFF}) for $s_{0}$ for any  values  $M,N$, because  the condition 2) imply that $[u_0, s_0]=0$.

Now put 
$$
v'_{m+1,0}:=v_{m+1} -(\sum v_{-1}\partial_2(s_{0k}u_0^k)+\sum [v_0,s_{0k}u_0^k]+ \sum (m+1)s_{0k}u_0^k\partial_2(v_{-1}))=
$$
$$
v_{m+1} -(\sum v_{-1}\partial_2(s_{0k})u_0^k+\sum 
(m+1)s_{0k}u_0^k\partial_2(v_{-1})),
$$
where the last equality follows from relation (\ref{tozhd}) applied to  $u_0^k$. For any $M,N$ we have 
\begin{equation}
\label{asdl}
F(M,N,v'_{m+1,0})\le F(M,N,v_{m+1}),
\end{equation}
where the equality holds only if  $F(M,N,v_{m+1})=0$. Indeed, this is true for  $v_{m+1} -\sum v_{-1}\partial_2(s_{0k})u_0^k$ because of the construction of $s_{0}$ and because $v_{-1}$ is a monic zeroth order operator. This is also true for $\sum (m+1)s_{0k}u_0^k\partial_2(v_{-1})$, because  $\ord_{\partial_1}((m+1)s_{0k}u_0^k\partial_2(v_{-1}))<k$ for all $k$.  Now it is only remain to apply the property ii) of the function $F$.

Put $u_{m,0}:=(m+1)s_0\partial_2(u_0)$. Again, since  $\ord_{\partial_1}(\partial_2(u_0))\le 0$ we have   
\begin{equation}
\label{asdl1}
F(M,N,u_{m,0})\le F(M,N,v_{m+1}). 
\end{equation}

Now define $d_0$ to be a solution of the equation $[u_0,d_0]=u_{m,0}$ satisfying the property 1) of the solution $s_0$. As we have seen above in the construction of such solutions, a solution with this property can be easily constructed. Since $\ord_{\partial_1}(v_{-1})=0$, we have
 for any $M,N$ 
\begin{equation}
\label{asdl2}
F(M,N,[v_{-1}, d_{0}])\le F(M,N,u_{m,0}) \le F(M,N,v_{m+1}),
\end{equation}
where the  equalities hold only if  $F(M,N,v_{m+1})=0$. 

Combining all these observations, we get for the element $v_{m+1,0}:=v'_{m+1,0}-[v_{-1}, d_{0}]$ and any $M,N$ 
\begin{equation}
\label{asdl3}
F(M,N,v_{m+1,0}) \le F(M,N,v_{m+1}),
\end{equation}
where the  equality holds only if  $F(M,N,v_{m+1})=0$. 

Moreover, as it follows from formulas  (\ref{formulka1}), (\ref{formulka2}), (\ref{formulka0}), $v_{m+1,0}$ is the $(m+1)$th coefficient of the operator $S_2^{-1}S_1^{-1}MS_1S_2$, where $S_1=1+s_0\partial_2^{m+1}$, $S_2=1+d_0\partial_2^{m}$. 
Using the relation from lemma \ref{commut} for operators $S_2^{-1}S_1^{-1}MS_1S_2$, $S_2^{-1}S_1^{-1}LS_1S_2$ for $\alpha =m+1$ we get $[u_0, v_{m+1,0}]=0$. 

Now, repeating all the above arguments for $v_{m+1,0}$ instead of $v_{m+1}$, we obtain new elements $s_1, u_{m,1}, d_1, v_{m+1,1}$ with the same properties, where $v_{m+1}$ is replaced by $v_{m+1,0}$. Continuing this line of reasoning we get a sequences of elements $v_{m+1,i}$, $s_i$, $u_{m,i}$, $d_i$. Note that the series  $\sum s_i$, $\sum u_{m,i}$,  $\sum d_i$ converge, because by (\ref{asdl}), (\ref{asdl1}), (\ref{asdl2}) for any $M,N$ there are only finite number of elements $v_{m+1,i}$, $s_i$, $u_{m,i}$, $d_i$ such that $F(M,N,(.) )>0$. Moreover, we have $\lim_i v_{m+1,i}=0$. Indeed, by the formula  (\ref{asdl3}) 
the function $F(M,N, v_{m+1,i})$ strictly decrease for any fixed $M,N$, so 
the sequences of $k$th coefficients of the operators $ v_{m+1,i}$ converges to zero for any $k$. 

Now we can define $s:=\sum s_i$, $d:=\sum d_i$. Then we have 
$$
S_1^{-1}LS_1=u_0+\sum u_{m,i}+(<m), 
$$
\begin{multline*}
S_1^{-1}MS_1=v_{-1}\partial_2+v_0+v_{m+1}-\\
(\sum_k v_{-1}\partial_2(\sum_is_{ik})u_0^k+\sum_k 
(m+1)(\sum_is_{ik})u_0^k\partial_2(v_{-1}))+(< m+1)
\end{multline*}
and  
$$
S_2^{-1}S_1^{-1}LS_1S_2=u_0+(<m), \mbox{\quad} S_2^{-1}S_1^{-1}MS_1S_2=v_{-1}\partial_2+v_0+(\lim_i v_{m+1,i})+(< m+1).
$$
This proves i) of the theorem. \\

ii) In order to get the second statement of the theorem it is enough to note that the conditions imply (in the notation of the proof of i)
$$
[S'S^{-1},u_0]=0, \mbox{\quad} [S'S^{-1}, v_{-1}\partial_2+v_0]=0.
$$
Since $u_0, v_{-1}$ are monic, we can represent the operator $S'S^{-1}$ as a series in $u_0, (v_{-1}\partial_2+v_0)$ with some coefficients from $R$. By lemma \ref{nichts}, all these coefficients must belong to $\ker\partial_1$ because of the first condition above. The second condition together with the  identity (\ref{tozhd}) then imply that all these coefficients belong also to $\ker\partial_2$, where from we obtain our statement. \\

iii) The third assertion is an easy consequence of the analogous theorem in one dimensional case. Namely, for the operator $(M\partial_2^{-1})_{+}L_+$, because of the solvability of the equation $\partial_1(x)=cx$, there exists a zeroth order invertible operator $\bar{S'}\in {\widehat E}_{\bar R}$ such that $(M\partial_2^{-1})_{+}L_+=\bar{S'}^{-1}\partial_1\bar{S'}$ (see, for example, Th.1 in \cite{Pa} or lemma 7.5 in \cite{Mu2}). The coefficients of this operator are defined modulo elements from $\ker\partial_1$. So, $v_{-1}:=\partial_1\bar{S'}L_+^{-1}\bar{S'}^{-1}=\partial_1u_0^{-1}$, where $u_0:=\bar{S'}^{-1}L_+\bar{S'}$ does not depend on the choice of $\bar{S'}$. 

Taking the decomposition of the element $(v_0)_-=(\tilde{v_0})_-+x$, where $x\in \ker\partial_1$ is a zeroth order operator and $(\tilde{v_0})_-\cap \ker\partial_1=0$, we can find a zeroth order invertible operator $\bar{S''}\in {\widehat E}_{\bar R}$, $\bar{S''}\in 1+{\widehat E}_{\bar R-}$, satisfying the equation $\partial_2(\log \bar{S''})=-x$ and belonging to $\ker\partial_1$. The coefficients of such operator will be defined modulo elements from $\ker\partial_1\cap \ker\partial_2=k[[\ldots ,t_{ij},\ldots ]]$. So, the operator $\bar{S}:=\bar{S'}\bar{S''}$ will satisfy the assertion and, as we have seen, it is defined modulo operators with coefficients from $k[[\ldots ,t_{ij},\ldots ]]$. \\
$\Box$ 

\begin{corol}
\label{corollary}
If the operators $L,M$ from the theorem belong to the ring $E_R$, then the operator $S$ can be found in the group $\Nu_R$.
\end{corol}

{\bf Proof.} To prove the corollary one can follow the proof of the theorem, (i) and note that the subsequent approximations of the operator $S$ belong, in fact, to the group $\Nu_R$. \\
$\Box$

{\bf Remark.} Combining the proof of proposition \ref{trivialitaet} with some  arguments from the proof of item i) of theorem \ref{Parshin}, one can easily prove that the system (\ref{Par}) has only trivial solutions of the form (\ref{L}), (\ref{M}) also in the ring ${\widehat E}_R$. Namely, one should understand $\ord_{\partial_1}(.)$ there as a function defined by the formula 
$$
\ord_{\partial_1}(a)=\min\{M| F(M,v_2(a),a)=0\}.
$$ 
With this definition and with lemma \ref{nichts} all arguments remain true. 

\bigskip

Now we can prove lemma \ref{3.1}.

{\bf Proof of lemma \ref{3.1}.} Let $P=\sum a_k\partial_2^k$, $a_k\in \widehat{E}_{\bar R}$. Since $P\in \widehat{E}_R$, its coefficients satisfy a growth order condition 
$$
v_2(a_k)>C_Pk \mbox{\quad for all } k>M_P.
$$
Let $Q=1-P=\sum b_k\partial_2^k\in \widehat{E}_R$. Since $(\pi (Q\partial_2^{-1}))_+=0$, $v_2(b_k)\ge 1$ for all $k\ge 1$. Of course the coefficients of $Q$ satisfy the same growth order condition 
$$
v_2(b_k)>C_Pk \mbox{\quad for all } k>M_P.
$$
Thus there exists a positive real number $J$ such that 
$$
v_2(b_k)\ge Jk \mbox{\quad for all } k\ge 1.
$$
Actually, $J=C_P/(1+C_pM_P)$ will do. Let $Q^n=\sum b_{n,k}\partial_2^k\in \widehat{E}_R$. Then we have 

{\it Claim.} i) For every $n,k\ge 1$ we have $v_2(b_{n,k})\ge Jk$. 

ii) The function $F(M,N,b_{n,k})$ satisfy the following property: for any given $M,N,k$ there exists a natural number $T(M,N,k)$ such that $F(M,N,b_{n,k})=0$ for all $n>T(M,N,k)$.

Here we take the function $F$ defined in step a) of the proof of theorem \ref{Parshin} with respect to $u_0=\partial_1$. 

{\it Proof of the claim.} 1) To prove the first assertion, we use an induction on $n$. Since $Q^{n+1}=Q\cdot Q^n$, by formula (\ref{star1}) of lemma \ref{kol'co2} we have 
\begin{equation}
\label{tex}
b_{n+1,k}=\sum_{q\in \sdz}\sum_{l\ge 0}C_q^lb_q\partial_2^l(b_{n,k+l-q})= \sum_{q=k}^{\infty}\sum_{l\ge 0}C_q^lb_q\partial_2^l(b_{n,k+l-q}) +
\end{equation}
$$
\sum_{q=0}^{k-1}\sum_{l\ge 0}C_q^lb_q\partial_2^l(b_{n,k+l-q}) + \sum_{q=-\infty}^{-1}\sum_{l\ge 0}C_q^lb_q\partial_2^l(b_{n,k+l-q}).
$$
In the first summation, we have $v_2(b_q\partial_2^l(b_{n,k+l-q}))\ge v_2(b_q)\ge Jq\ge Jk$ for all $q\ge k\ge 1$. In the second summation,  since $v_2(\partial_2^l(b_{n,i}))\ge v_2(b_{n,i})$, by the induction hypothesis we have 
$$
v_2(b_q\partial_2^l(b_{n,k+l-q}))\ge v_2(b_q)+ v_2(b_{n,k+l-q})\ge Jq+J(k+l-q)\ge Jk
$$
for all $0\le q\le k-1$. In the third summation we have $v_2(b_q\partial_2^l(b_{n,k+l-q}))\ge v_2( b_{n,k+l-q})\ge Jk$ for all $k\le -1$. Hence $v_2(b_{n+1,k})\ge Jk$. 

2) To prove the second assertion we use an induction on $N$ and $|k|$. First let's introduce a function $q:\dn\times\widehat{E}_{\bar R} \rightarrow \dz\cup\{-\infty\}$:
$$
\mbox{$q(N,a)$ {\it is the maximal number such that the coefficient $b_{q(N,a)}$ of $a$}}
$$
\begin{equation}
\label{tex1}
\mbox{\it{ satisfy the property $v_2(b_{q(N,a)})\le N$;}}
\end{equation}
$$
\mbox{\it{ if there are no such coefficients, we put $q(N,a)=-\infty$}}
$$
Using formula (\ref{star}) of lemma \ref{kol'co1} it is easy to obtain the following property:
\begin{equation}
\label{tex2}
q(N,ab)\le q(N,a)+q(N,b).
\end{equation}
If $v_2(a)>0$ and $v_2(b)>0$, we also have 
\begin{equation}
\label{tex3}
q(N,ab)\le q(N-1,a)+q(N-1,b).
\end{equation}
Indeed, in this case by (\ref{star}) we have: $v_2((ab)_k)>N$ for $k>q(N-1,a)+q(N-1,b)$ if $v_2(a_i)>N-1$ for $i>q(N-1,a)$ and $v_2(b_i)>N-1$ for $i>q(N-1,b)$. 

Analogously, if $v_2(a)>0$ and $v_2(b)=0$, we have 
\begin{equation}
\label{tex31}
q(N,ab)\le q(N,a)+q(N-1,b),
\end{equation}
and if $v_2(a)=0$ and $v_2(b)>0$
\begin{equation}
\label{tex32}
q(N,ab)\le q(N-1,a)+q(N,b).
\end{equation}

3) Note that it suffice to prove our assertion only for $M<0$. For such $M$ we have then the following corollaries of properties (\ref{tex2}), (\ref{tex3}):
\begin{equation}
\label{tex4}
\mbox{$F(M,N,ab)=0$ if $F(M,N,a)=F(M,N,b)=0$;  } 
\end{equation}
and if $v_2(a)>0$ and $v_2(b)>0$, we also have
\begin{equation}
\label{tex5}
\mbox{$F(M,N,ab)=0$ if $F(M,N-1,a)=F(M,N-1,b)=0$.  } 
\end{equation}
Indeed, the condition $F(M,N,a)=F(M,N,b)=0$ means $M\ge q(N,a)$, $M\ge q(N,b)$. Therefore, by (\ref{tex2}) $M>2M\ge q(N,a)+q(N,b)\ge q(N,ab)$, where from $F(M,N,ab)=0$. Analogously, the condition $F(M,N-1,a)=F(M,N-1,b)=0$ together with (\ref{tex3}) imply $F(M,N,ab)=0$. 

Obviously, by (\ref{tex31}) we also have for $v_2(a)>0$ and $v_2(b)=0$
\begin{equation}
\label{tex51}
\mbox{$F(M,N,ab)=0$ if $F(M,N,a)=F(M,N-1,b)=0$,  } 
\end{equation}
and by (\ref{tex32}) we have for $v_2(a)=0$ and $v_2(b)>0$
\begin{equation}
\label{tex52}
\mbox{$F(M,N,ab)=0$ if $F(M,N-1,a)=F(M,N,b)=0$.  } 
\end{equation}

4) Now let's prove the first step of our double induction. Let $N=0$ and $|k|=0$. Then we have $T(M,N,k)=|M|$ and $q(N,b_{n,k})\le -n$. Indeed, by (\ref{tex}) 
$$
F(M,0,b_{n+1,0})=F(M,0,b_0b_{n,0})=\ldots =F(M,0,b_0^{n+1}),
$$ 
because $v_2(b_{j,l})\ge Jl>0$ for all $j,l\ge 1$ by the first assertion of the claim, so all the summands in (\ref{tex}) except $b_0b_{n,0}$ have valuation greater than zero and therefore they don't change the value of the function $F(M,0,.)$. Obviously, the same is true for the function $q(0,.)$. Since $\ord_{\partial_1}b_0\le -1$, we have $\ord_{\partial_1}b_0^n\le -n$, where from $F(M,0,b_0^{n+1})=0$ for all $n>|M|$ and $q(0,b_{n+1,k})\le -n-1$. 

5) Now let $N=0$ and $|k|>0$. Since $v_2(b_{j,l})>0$ for all $j,l\ge 1$, we have $F(M,0, b_{n,k})=0$, $q(0,b_{n,k})=-\infty$ for all $n$ and $k\ge 1$. So, we can assume $k<0$. By the induction hypothesis $F(M,0,b_{n,j})=0$ for all $n>T(M,0,j)$ if $|j|<|k|$. Denote by $N_0$ the maximum of all such $T(M,0,j)$. Now consider the elements $b_{n+N_0,k}$ with $n>N_0$. 
Since $Q^{n+N_0}=Q^nQ^{N_0}$, we have by the same reason as above\\
$F(M,0,b_{n+N_0,k})=$
\begin{equation}
\label{tex6}
F(M,0,\sum_{i=k}^0\sum_{l=0}^{-k}C_i^lb_{n,i}\partial_2^l(b_{N_0,k+l-i}))\le \max_{k\le i\le 0, 0\le l\le -k}\{F(M,0,b_{n,i}\partial_2^l(b_{N_0,k+l-i}))\},
\end{equation}
where the last inequality follows from the property ii) of the function $F$, see step a) of theorem \ref{Parshin}.

6) Now it suffice to prove the assertion for each function $F(M,0,b_{n,i}\partial_2^l(b_{N_0,k+l-i}))$, where $k\le i\le 0, 0\le l\le -k$ and $k+l-i\le 0$, and then define the number $T(M,0,k)$ as a maximum of corresponding numbers for each pair $(l,i)$. 

The set of all pairs $(l,i)$ can be divided in two subsets: $\{(0,0),(0,k)\}$ and all other pairs. For the pairs from the second subset we have $|i|<|k|$ and $|k+l-i|<|k|$. So, for all these pairs we have $F(M,0,b_{n,i})=0$ and $F(M,0,\partial_2^l(b_{N_0,k+l-i}))\le F(M,0,b_{N_0,k+l-i})=0$ (the last inequality holds by the analog of the property i) of the function $F$ in step a) of the proof of theorem \ref{Parshin}). Therefore, by (\ref{tex4}) we obtain $F(M,0,b_{n,i}\partial_2^l(b_{N_0,k+l-i}))=0$ for all $n>N_0$ and all pairs from the second subset. 

Now consider the pair $(0,0)$ from the first subset. 
By the property (\ref{tex2}) we have 
$$
q(0,b_{n,0}\partial_2^l(b_{N_0,k}))\le q(0,b_{n,0})+q(0,\partial_2^l(b_{N_0,k}))\le q(0,b_{n,0})+q(0,b_{N_0,k})\le -n+q(0,b_{N_0,k}).
$$
Therefore, for all $n>-M+q(0,b_{N_0,k})$ we have $F(M,0,b_{n,0}\partial_2^l(b_{N_0,k}))=0$. 

7) Combining all together, we get for all $n>\max\{N_0,-M+q(0,b_{N_0,k})\}:=\tilde{T}$ (see (\ref{tex6})):
\begin{equation}
\label{tex7}
F(M,0,b_{n+N_0,k})\le F(M,0,b_{n,k}\partial_2^l(b_{N_0,0})).
\end{equation}

Now put 
$$O=(\max_{\tilde{T}\le j\le \tilde{T}+N_0}\{F(M,0,b_{j,k})\})$$ 
and put $T(M,0,k)=\tilde{T}+ON_0$. We claim, that for all $n>T(M,0,k)$ $F(M,0,b_{n,k})=0$. 

Assume the converse. Let $n>T(M,0,k)$ be the number such that $F(M,0,b_{n,k})>0$. Then by (\ref{tex7}) and (\ref{tex2}) we have 
$$
0<F(M,0,b_{n,k})=q(0,b_{n,k})-M\le F(M,0,b_{n-N_0,k}\partial_2^l(b_{N_0,0}))=q(0,b_{n-N_0,k}\partial_2^l(b_{N_0,0}))-M
$$
$$
\le q(0,b_{n-N_0,k})+q(0,b_{N_0,0})-M\le q(0,b_{n-N_0,k})-M -N_0=F(M,0,b_{n-N_0,k})-N_0\le \ldots 
$$
$$
\le F(M,0,b_{n-\tilde{O}N_0,k})-\tilde{O}N_0,
$$
where $\tilde{T}\le n-\tilde{O}N_0\le \tilde{T}+N_0$ and therefore $\tilde{O}>O$. So, 
$$
F(M,0,b_{n-\tilde{O}N_0,k})-\tilde{O}N_0\le O-\tilde{O}N_0\le 0,
$$
a contradiction. 

8) Now assume $N$ is an arbitrary positive number. Let $k_0:=[N/J]$ be the
integral part of the number $N/J$. For all $k>k_0$ we have $v_2(b_{n,k})\ge Jk>N$, so the assertion is trivial for all such $k$. 

To prove the assertion we will use an inverse induction on $k\le k_0$. 
By the induction hypothesis on $N$ there exists a natural number $N_0$ such that $F(M,N-1,b_{n,i})=0$ for all $n\ge N_0$ and $k_0\le i\le 0$ (with fixed $M$). We then have for all $n>N_0$ and $k=k_0$ 
$$
F(M,N,b_{n+N_0,k_0})=
F(M,N,\sum_{i=0}^{k_0}\sum_{l=0}^{k_0}C_i^lb_{n,i}\partial_2^l(b_{N_0,k_0+l-i}))\le 
$$
$$
\max_{k_0\ge i\ge 0, 0\le l\le k_0}\{F(M,N,b_{n,i}\partial_2^l(b_{N_0,k_0+l-i}))\}\le 
$$
$$
\max\{F(M,N,b_{n,0}\partial_2^l(b_{N_0,k_0})), F(M,N,b_{n,k_0}\partial_2^l(b_{N_0,0}))\},
$$
where the last inequality follows from (\ref{tex5}). For all $n>-M+q(N,b_{N_0,k_0})$ we have, as above, $F(M,N,b_{n,0}\partial_2^l(b_{N_0,k_0}))=0$. So, for all sufficiently large $n$ 
$$
F(M,N,b_{n+N_0,k_0})\le F(M,N,b_{n,k_0}\partial_2^l(b_{N_0,0})),
$$
and we can repeat the arguments of step 7) to get the proof of the assertion in the case $k=k_0$. 

9) Let's prove the assertion for arbitrary $k<k_0$. By the induction hypothesis on $N, k$ there exists a natural number $N_0$ such that $F(M,N-1,b_{n,j})=0$ for all $n\ge N_0$ and $k_0\ge j\ge k-k_0$, and $F(M,N,b_{n,j})=0$ for all $n\ge N_0$ and $k_0\ge j>k$ (again $M$ is  fixed). We then have for all $n>N_0$  
$$
F(M,N,b_{n+N_0,k})=
F(M,N,\sum_{i=k-k_0}^{k_0}\sum_{l=0}^{2k_0-k}C_i^lb_{n,i}\partial_2^l(b_{N_0,k+l-i}))\le 
$$
$$
\max_{k_0\ge i\ge (k-k_0), 0\le l\le 2k_0-k}\{F(M,N,b_{n,i}\partial_2^l(b_{N_0,k+l-i}))\}\le 
$$
$$
\max\{F(M,N,b_{n,0}\partial_2^l(b_{N_0,k})), F(M,N,b_{n,k}\partial_2^l(b_{N_0,0}))\},
$$
where the last inequality follows from (\ref{tex4}), (\ref{tex5}), (\ref{tex51}), (\ref{tex52}). Repeating the arguments of the end of step 8), we obtain the assertion also in this case, hence in general. \\
The claim is proved.
\smallskip

By the claim we can conclude that $\sum_{n=0}^{\infty}b_{n,k}\in \widehat{E}_{\bar R}$ is a well-defined element for any $k$. Therefore
$$
\sum_{n=0}^{\infty}Q^n=1+\sum_{n=1}^{\infty}\sum_{k\in\sdz}b_{n,k}\partial_2^k=1+\sum_{k\in\sdz}(\sum_{n=1}^{\infty}b_{n,k})\partial_2^k.
$$
By the first assertion of the claim $v_2(\sum_{n=1}^{\infty}b_{n,k})\ge Jk$ if $k>0$. Obviously, we also have $b_{n,0}|_{t=0}=0$ for all $n\ge 1$, where from $\sum_{n=1}^{\infty}b_{n,0})|_{t=0}=0$. This implies that $\sum_{n=0}^{\infty}Q^n\in \widehat{E}_{R}^{\times}$. Since 
$$
1=\sum_{n=0}^{\infty}(1-P)^n-\sum_{n=1}^{\infty}(1-P)^n=\sum_{n=0}^{\infty}(1-P)^n-(1-P)\sum_{n=0}^{\infty}(1-P)^n=P\sum_{n=0}^{\infty}(1-P)^n,
$$
the Neumann series gives $P^{-1}\in \widehat{E}_{R}^{\times}$.\\
The lemma is proved.\\
$\Box$

\section{Modified Sato-Wilson systems}

\subsection{Equivalence of modified KP and modified Sato-Wilson systems}

In this subsection we will prove the equivalence of systems $(KP)_{\alpha}$ and appropriately generalized Sato-Wilson systems. This is a generalization of well known equivalence from the classical case. It is also the necessary step in solving the Cauchy problem for the systems $(KP)_{\alpha}$.

Let's fix a system $(KP)_{\alpha}$. 
For a given operator $N=(L,M)$ we define formal 1-forms by 
$$
Z^{\alpha}_{\pm}=\pm \sum_{i\in \sdz ,j\in \sdz_+, i\le \alpha j}dt_{i,j}(L^iM^j)_{\pm}
$$
Then it is easy to see that the Lax equation for the $(KP)_{\alpha}$ system is given by 
\begin{equation}
\label{ZZ}
d_{\alpha}N=[Z_{+}^{\alpha},N]=[Z_{-}^{\alpha},N],
\end{equation}
where $d_{\alpha}N$ denotes the vector $(d_{\alpha}L, d_{\alpha}M)$, $[Z_{\pm}^{\alpha},N]$ denotes  the vector 
$([Z_{\pm}^{\alpha},L], [Z_{\pm}^{\alpha},M])$, and 
$d_{\alpha}=\sum_{i\in \sdz , j\in \sdz_+, i\le \alpha j}dt_{i,j}\partial /\partial t_{i,j}$ denotes the exterior derivative in $t$. 

\begin{prop}
\label{svyaznost'}
The Lax equation for the  $(KP)_{\alpha}$ system is equivalent to the integrability condition 
$$
d_{\alpha}Z_{\pm}^{\alpha}=Z_{\pm}^{\alpha}\wedge Z_{\pm}^{\alpha}
$$
\end{prop}

{\bf Proof.} In \cite{Pa}, prop. 4, i) one can find the proof that the solution of the $(KP)_{\alpha}$ system in the form (\ref{L}), (\ref{M})  also satisfies the system 
\begin{equation}
\label{zahar}
\frac{\partial (L^iM^j)_{+}}{\partial t_{l,m}}-
\frac{\partial (L^lM^m)_{+}}{\partial t_{i,j}}=[(L^lM^m)_{+},(L^iM^j)_{+}],
\end{equation}
where $i,l\in\dz$, $j,m\in\dz_+$, $i\le \alpha j$, $l\le \alpha m$, 
that is, to the equation $dZ_{+}^{\alpha}=Z_{+}^{\alpha}\wedge Z_{+}^{\alpha}$. Since for $Z^{\alpha}=Z_{+}^{\alpha}-Z_{-}^{\alpha}=\sum dt_{i,j}L^iM^j$ we have $Z^{\alpha}\wedge Z^{\alpha}=0=Z_{+}^{\alpha}\wedge Z_{+}^{\alpha}+Z_{-}^{\alpha}\wedge Z_{-}^{\alpha}$, we have also $dZ_{-}^{\alpha}=Z_{-}^{\alpha}\wedge Z_{-}^{\alpha}$. 

To prove the converse assertion let us consider first the equations (\ref{zahar}) with arbitrary $l,m,j$ and with $i=-|\alpha j|$. Clearly, $i\le \alpha j$ for any $j$. The equations (\ref{zahar}) can be rewritten as 
$$
\frac{\partial (L^iM^j)}{\partial t_{l,m}}-[(L^lM^m)_{+},(L^iM^j)]=-[(L^lM^m)_{+},(L^iM^j)_{-}]+\frac{\partial (L^lM^m)_{+}}{\partial t_{i,j}}+\frac{\partial (L^iM^j)_{-}}{\partial t_{l,m}},
$$
where from  
$$
\ord_{\partial_2}
\left (
\frac{\partial (L^iM^j)}{\partial t_{l,m}}-[(L^lM^m)_{+},L^iM^j]
\right )
\le m 
$$ 
Since $\partial /\partial t_{l,m}$ and $[(L^lM^m)_{+},]$ are derivations, we have 
$$
\left (
\frac{\partial}{\partial t_{l,m}}-[(L^lM^m)_{+},]
\right )
L^iM^j=\left ( \sum_{k=i+1}^{0}L^k\left ( 
\frac{\partial L^{-1}}{\partial t_{l,m}}-[(L^lM^m)_{+},L^{-1}]\right ) L^{i+1-k}
\right ) M^j +
$$
$$
L^i\sum_{k=0}^{j-1} M^k \left (
\frac{\partial M}{\partial t_{l,m}}-[(L^lM^m)_{+},M]\right ) M^{j-1-k},
$$
where from 
$$
\ord_{\partial_2}\left ( 
\sum_{k=i+1}^{0}L^k\left ( 
\frac{\partial L^{-1}}{\partial t_{l,m}}-[(L^lM^m)_{+},L^{-1}]\right ) L^{i+1-k}
 M + \right .
$$
\begin{equation}
\label{zah}
\left .
L^i\sum_{k=0}^{j-1} M^k \left (
\frac{\partial M}{\partial t_{l,m}}-[(L^lM^m)_{+},M]\right ) M^{-k}\right )
\le m-j+1
\end{equation}

Obviously, if $(L,M)$ are solutions of (\ref{zahar}) in the form (\ref{L}), (\ref{M}), then $u_0, v_{-1}$ do not depend on times and therefore $L,M$ are invertible. Hence, since the characteristic of the  ground field $k$ is zero, we have 
\begin{equation}
\label{zah1}
\ord_{\partial_2}
\left (
L^i\sum_{k=0}^{j-1} M^k \left (
\frac{\partial M}{\partial t_{l,m}}-[(L^lM^m)_{+},M]\right ) M^{-k}
\right ) = \ord_{\partial_2}\left (
\frac{\partial M}{\partial t_{l,m}}-[(L^lM^m)_{+},M]\right ) 
\end{equation}
and
\begin{equation}
\label{zah2}
\ord_{\partial_2}\left (
\sum_{k=i+1}^{0}L^k\left ( 
\frac{\partial L^{-1}}{\partial t_{l,m}}-[(L^lM^m)_{+},L^{-1}]\right ) L^{i+1-k}
\right )=\ord_{\partial_2}\left ( 
\frac{\partial L^{-1}}{\partial t_{l,m}}-[(L^lM^m)_{+},L^{-1}]\right ).
\end{equation}
Let's denote by $Q_L$ the coefficient of the series $W_1=\frac{\partial L^{-1}}{\partial t_{l,m}}-[(L^lM^m)_{+},L^{-1}]$ by the power $\partial_1^{q(v_2(W_1),W_1)}\partial_2^{\ord_{\partial_2}(W_1)}$,  and by 
$Q_M$ the coefficient of the series $W_2=\frac{\partial M}{\partial t_{l,m}}-[(L^lM^m)_{+},M]$ by the power $\partial_1^{q(v_2(W_2),W_2)}\partial_2^{\ord_{\partial_2}(W_2)}$ (where the function $q$ was defined in the proof of lemma \ref{3.1}, and we assume that $W_1,W_2\ne 0$). Let $Q'_L$, $Q'_M$ be analogous coefficients of the series from the left hand sides of (\ref{zah2}), (\ref{zah1}). By the definition we have $v_2(Q'_L)=v_2(Q_L)=v_2(W_1)$, $v_2(Q'_M)=v_2(Q_M)=v_2(W_2)$. 
Then the equality of orders in (\ref{zah2}), (\ref{zah1}) exactly means that we have 
$$
v_2(Q'_L-iQ_L)>v_2(W_1), \mbox{\quad} v_2(Q'_M-jQ_M)>v_2(W_2).
$$
Since $j$ can be an arbitrary large number and the orders 
$$
\ord_{\partial_2}\left (
\frac{\partial M}{\partial t_{l,m}}-[(L^lM^m)_{+},M]\right ) , \mbox{\quad} \ord_{\partial_2}\left ( 
\frac{\partial L^{-1}}{\partial t_{l,m}}-[(L^lM^m)_{+},L^{-1}]\right )
$$
don't depend on $j$, we obtain from (\ref{zah})
$$
\ord_{\partial_2}\left (
\frac{\partial M}{\partial t_{l,m}}-[(L^lM^m)_{+},M]\right )=\ord_{\partial_2}
\left (
\frac{\partial L^{-1}}{\partial t_{l,m}}-[(L^lM^m)_{+},L^{-1}]
\right ) +1
$$
and since $L,M$ are monic,
$$
v_2(W_1)=v_2(W_2), \mbox{\quad} v_2(Q'_M-Q'_L)>v_2(W_1),
$$
where from 
$$
v_2(iQ_L-jQ_M)>v_2(W_1)
$$
for all $j$ (recall that $i$ depend on $j$). The same arguments with $i'=-2|\alpha j|$ give us 
$$
v_2(i'Q_L-jQ_M)=v_2(2iQ_L-jQ_M)>v_2(W_1), 
$$
where from $v_2(W_1)<v_2(2iQ_L-iQ_L)=v_2(iQ_L)=v_2(W_1)$, a contradiction. Therefore, our assumption about $W_1,W_2$ is wrong, i.e. $W_1=W_2=0$. But this means that the system (\ref{ZZ}) holds.\\
The proposition is proved.\\
$\Box$

\begin{corol}
\label{ochev}
The equations (\ref{ZZ}) are compatible, i.e. $d_{\alpha}[Z^{\alpha}_{+},N]=0$.
\end{corol}

The proof is clear. \\

Suppose that our $(KP)_{\alpha}$ system with initial condition $N_0=(L_0,M_0)$ has a solution $N\in {\widehat{E}}_R$. Then by theorem \ref{Parshin}, (i) there exists $S\in \widehat{\Nu}_R$ with $L=Su_0S^{-1}$, $M= S(v_{-1}\partial_2+v_0)S^{-1}$. As we have already seen, $u_0, v_{-1}$ do not depend on times and therefore coinside with the first coefficients of the initial condition $N_0$. Notably, $v_0$ constructed in the theorem, also don't depend on times. Indeed, if $v'_0$ is the coefficient of $M$ as in (\ref{dva}), we have 
$$
\frac{\partial v'_0}{\partial t_{ij}}=-[(L^iM^j)_-,M]_0=-[x,v_{-1}],
$$
where $x$ is the first coefficient of $(L^iM^j)_-$ represented as series in $\partial_2^{-1}$. For $x$ we have also another equation: 
\begin{equation}
\label{nord}
\frac{\partial u_1}{\partial t_{ij}}=-[(L^iM^j)_-,L]_1=-[x,u_0].
\end{equation}
Therefore, from definition of $v_0$ follows 
$$
\frac{\partial v_0}{\partial t_{ij}}=\frac{\partial v'_0}{\partial t_{ij}}-[v_{-1},\frac{\partial b_{u_1}}{\partial t_{ij}}]=0,
$$
because $\frac{\partial b_{u_1}}{\partial t_{ij}}$ can be an arbitrary solution of the equation (\ref{nord}). 

So, denote $L_+=u_0$ by $L_{00}$ and $(v_{-1}\partial_2+v_0)$ by $M_{00}$.  As we have  shown above, $L_{00},M_{00}$ are uniquely defined  by $N_0$. They are also invertible because of our assumptions on the initial condition. 
Now let 
\begin{equation}
\label{uni}
\omega_{\alpha}^{N_0} =\sum_{i\in\sdz ,j\in \sdz_+, i\le \alpha j}dt_{i,j}L_{00}^iM_{00}^j 
\end{equation}
It's clear that $S\omega_{\alpha}^{N_0} S^{-1}=\sum dt_{i,j}L^jM^j=Z^{\alpha}$. Let's introduce the additional notation:
$$
E_{k,L_{00},M_{00}}:=k((L_{00}^{-1}))((M_{00}^{-1})), \mbox{\quad} 
\Nu_{k,L_{00},M_{00}}:=\{1+E_{k,L_{00},M_{00},-}\} ,
$$
$$
\widehat{\Nu}_{k,L_{00},M_{00}}:=\{L=1+\sum_{i=1}^{\infty}a_iM_{00}^{-i}| a_i=\sum_{q\in\sdz}b_qL_{00}^q\in \widehat{E}_{\bar{R}} \mbox{\quad {\it and} $b_q\in k[[\ldots ,t_{ij},\ldots ]]$}\}.
$$

\bigskip

From now on {\it we will make one more assumption}. Since we have fixed the system $(KP)_{\alpha}$ and this system contains no times $t_{ij}$ with $i>\alpha j$ and since in general $(\partial /\partial t_{kl}N)(t_{ij}=0)=\partial /\partial t_{kl}(N(t_{ij}=0))$ for $i>\alpha j$ and $k\le \alpha l$ and since $[(L^kM^l)_+,N](t_{ij}=0)=[(L(t_{ij}=0)^kM(t_{ij}=0)^l)_+,N(t_{ij}=0)]$, we can {\it assume that solutions} (if they exist) {\it of the system do not depend on times} $t_{ij}$ {\it with} $i>\alpha j$. Obviously, {\it we can also assume that solutions do not depend on} $t_{00}$. 

\bigskip

\begin{prop}
\label{SW}
The system $(KP)_{\alpha}$ with the initial condition $N_0$ is equivalent to the modified Sato-Wilson system: 
$$
d_{\alpha}S=-(S\omega_{\alpha}^{N_0}S^{-1})_-S \eqno{(SW)_{\alpha}}
$$
with $S\in \widehat{\Nu}_R\cdot\widehat{\Nu}_{k,L_{00},M_{00}}$ and with initial condition $S(0)\in (1+E_-)\cdot \Nu_{k,L_{00},M_{00}}$.
\end{prop}

{\bf Proof.} If $S$ is a solution of the system $(SW)_{\alpha}$, then $N=(SL_{00}S^{-1},SM_{00}S^{-1})$ gives a solution of the $(KP)_{\alpha}$, the proof is the same as in \cite{Pa}, prop.4, ii). 

Let's prove the converse. Let $N$ be a solution of the $(KP)_{\alpha}$, and let 
$$N=(S_0L_{00}S_0^{-1}, S_0M_{00}S_0^{-1})$$
for some $S_0$.  Put $Z^0_{-}=S_0^{-1}Z^{\alpha}_{-}S_0-
S_0^{-1}d_{\alpha}S_0$. Then we have $d_{\alpha}Z^0_{-}=Z^0_{-}\wedge Z^0_{-}$. This follows from the same argument as in the classical case of the one dimensional $(KP)$ system, namely, because $Z^0_{-}$ is a gauge transformation of the flat connection $Z^{\alpha}_{-}$. We regard here $Z^{\alpha}_{\pm}$ as connections on the trivial bundle $E\times T$, where $T$ is the space of deformation parameters $t=(t_{ij})$, on which the Lie algebra $E$ acts by the commutator.
One can check this also directly.  

Now let us show that all operators in $Z^0_{-}$ (written as series in $L_{00}^{-1},M_{00}^{-1}$) have only coefficients belonging to $k[[\ldots ,t_{ij}, \ldots ]]$:
$$
[Z^0_{-},L_{00}]=S_0^{-1}[S_0Z^0_{-}S_0^{-1},S_0L_{00}S_0^{-1}]S_0=
S_0^{-1}[Z^{\alpha}_{-}-d_{\alpha}S_0S_0^{-1},L]S_0=
$$
$$
S_0^{-1}([Z^{\alpha}_{-},L]-[dS_0S_0^{-1},L])S_0=
S_0^{-1}(d_{\alpha}L-d_{\alpha}L)S_0=0
$$ 
(and analogously for $M_{00}$), because 
$$
d_{\alpha}L=d_{\alpha}S_0L_{00}S_0^{-1}-S_0L_{00}S_0^{-1}d_{\alpha}S_0S_0^{-1}
=[d_{\alpha}S_0S_0^{-1},L].
$$
Now consider the equation $d_{\alpha}C=Z^0_{-}C$. Since $d_{\alpha}Z^0_{-}=Z^0_{-}\wedge Z^0_{-}$, this equation is compatible and therefore there exists a solution $C\in \widehat{\Nu}_{k,L_{00},M_{00}}$. 
Let's explain this fact, because here is important the choice of the topology in the ring $\bar R$ and our assumptions on the solutions of the system $(KP)_{\alpha}$. 

To prove the existence of a solution, let's fix some bijection between the set of all pairs $(i,j)$ with $i\in\dz$, $j\in\dz_+$, $i\le \alpha j$ and natural numbers, $f:\{(i,j)\}\rightarrow \dn$, such that for any finite set of integers $(a_1,\ldots a_n)$ there exists $J\in \dn$ with the following property: for all $k>J$ $f^{-1}(k)=(i_k,j_k)$, where either $j_k>n$ or $i_k<a_{j_k}$. Such a bijection can be constructed, for example, by counting integral points (lying below the graphic of the function $\alpha$) lying inside  polygons formed by parallel lines that go through the points $(i,0)$, $(i-1,0)$, $i\le \alpha (0)$, 
and are parallel to the line going through the points $(\alpha (0)-1,0)$ and $(\alpha (1), 1)$, the vertical lines $(.,k)$ and lines going through the points $(\alpha (k),k)$, $(\alpha (k+1),k+1)$.

Now $Z_-^0=\sum_{k=1}^{\infty}P_{f^{-1}(k)}dt_{f^{-1}(k)}$, where $\ord_{\partial_2}(P_{ij})\le -1$, and the solution $C=\exp (C')$, where $C'$ is a solution of the equation $d_{\alpha}C'=Z^0_-$. To find a solution $C'$ it suffice to find a solution $C''$ of the equation $d_{\alpha}C''= \sum_{k=2}^{\infty}P'_{f^{-1}(k)}dt_{f^{-1}(k)}$, where $P'_{f^{-1}(k)}=P_{f^{-1}(k)}-\partial /\partial t_{f^{-1}(k)}(\int P_{f^{-1}(1)}dt_{f^{-1}(1)})$, and then put $C'=C''+\int P_{f^{-1}(1)}dt_{f^{-1}(1)}$. Because of compatibility conditions on $Z^0_-$, $P'_{f^{-1}(k)}$ don't depend on  $t_{f^{-1}(1)}$ for all $k\ge 2$. Continuing this line of reasoning, we obtain that $C'$ must be equal to a sum of infinite number of elements from the ring $\widehat{E}_R$, where each summand is an integral on time and only finite number of summands depend on $t_{f^{-1}(k)}$ for any fixed $k$. 

This sum converges if and only if the infinite sum of elements from the ring $\bar R$ with the same property converges in $\bar R$. By definition of the ring $\bar R$ a series converges if and only if for any neighborhood of zero there are only finite number of monomials of all summands not belonging to the neighborhood.  By our assumption on the solutions of the system $(KP)_{\alpha}$ all summands don't depend on times $t_{ij}$ with $i>\alpha j$. From the other hand side, since all summands are integrals on times, there are no free terms, i.e. all monomials depend on some other times $t_{ij}$. From the definition of the base of neighborhoods  of zero in the ring $A_t$ follows that each monomial that don't depend on $t_{ij}$ with $i>\alpha j$ and don't belong to a neighborhood of zero, must depend on some $t_{ij}$ from some {\it finite} set of times $\{t_{kl}\}$.  It's easy to see that for each neighborhood such a finite set lies in the set $\{t_{f^{-1}(k)}, k=1,\ldots ,m\}$ for some finite number $m$. Since in our sum there are only  finite number of summands depending on finite number of such times and since all summands have no free terms, we obtain the convergence. 

So, the solution $C'$ exists. Since $\ord_{\partial_2}P_{f^{-1}(k)}\le -1$ for all $k$, we obtain  $\ord_{\partial_2}(C')\le -1$. Therefore $C:=\exp (C')$ also exists. 

Now put $S=S_0C$, then 
$$
d_{\alpha}S-Z^{\alpha}_{-}S=d_{\alpha}S_0C+S_0d_{\alpha}C-Z^{\alpha}_{-}S_0C=S_0(S_0^{-1}d_{\alpha}S_0+d_{\alpha}CC^{-1}-S_0^{-1}Z^{\alpha}_{-}S_0)C=0
$$
and $SL_{00}S^{-1}=S_0L_{00}S_0^{-1}=L$, $SM_{00}S^{-1}=S_0M_{00}S_0^{-1}=M$.

At last, as it follows from corollary \ref{corollary}, the initial condition $S(0)\in (1+E_-)\cdot \Nu_{k,L_{00},M_{00}}$.
$\Box$

{\bf Remark.} As it follows from the proof and corollary \ref{corollary}, if $N\in E_R$, then $S\in \Nu_R\cdot \widehat{\Nu}_{k,L_{00},M_{00}}$, and vice versa.

\begin{prop}
\label{bijection}
We have a bijection between the following two sets:
\begin{multline*}
Sol((KP)_{\alpha})=\{ N=(L,M),\mbox{\quad} L,M\in \widehat{E}_R (E_R) \mbox{\quad}|\mbox{\quad} N \mbox{\quad satisfies $(KP)_{\alpha}$ }\\
\mbox{ and  $L, M$  don't depend on $t_{ij}$, $i>\alpha j$} \}
\end{multline*}
and
\begin{multline*}
Sol((SW)_{\alpha})=\{S\in \widehat{\Nu}_R\cdot \widehat{\Nu}_{k,L_{00},M_{00}} ({\Nu}_R\cdot \widehat{\Nu}_{k,L_{00},M_{00}}) \mbox{\quad}|\mbox{\quad}
S \mbox{\quad satisfies $(SW)_{\alpha}$ }\\
\mbox{ and  does not  depend on $t_{i,j}$, $i>\alpha j$} \}/(\Nu_{k,L_{00},M_{00}})
\end{multline*}

\end{prop}

{\bf Proof.} The bijection is given by the formula $L=SL_{00}S^{-1}$, $M=SM_{00}S^{-1}$ and follows from the proposition \ref{SW}. Indeed, 
suppose $S_1N_{00}S_1^{-1}=S_2N_{00}S_2^{-1}$ for $N_{00}=(L_{00},M_{00})$. Then $S_2=S_1C$, where $C\in \widehat{\Nu}_{k,L_{00},M_{00}}$. Therefore, from the equation $(SW)_{\alpha}$ we get 
$$
Z_{-}^{\alpha}(S_1C)=d_{\alpha}(S_1C)=d_{\alpha}S_1C+S_1d_{\alpha}C, 
$$
where from $S_1d_{\alpha}C=0$. Since $S_1,S_2$ don't depend on $t_{i,j}$, $i>\alpha j$, we conclude $C\in {\Nu}_{k,L_{00},M_{00}}$.\\
The proposition is proved.\\
$\Box$

To complete the picture, we have to 
describe the set of all such initial conditions $S(0)\in (1+E_{-})$ that give us trivial initial conditions 
(so, trivial solutions by theorem \ref{1.4} below) of the system $(KP)_{\alpha}$. Following Mulase, \cite{Mu2}, we give the following definition. 

\begin{defin}

\label{adm}
A pseudo-differential operator $T\in E$ is said to be $(L_{00},M_{00})$-admissible if it is an invertible operator of order zero such that $TL_{00}T^{-1}, TM_{00}T^{-1}\in E_{k,L_{00},M_{00}}$. The set of all $(L_{00},M_{00})$-admissible operators is denoted by $\Gamma_{a,L_{00},M_{00}}$. 
\end{defin}

\begin{lemma}
\label{admissible}
Every $(L_{00},M_{00})$-admissible operator $T$ has the following form: 
$$
T=S^{-1}T_{0}a,
$$
$$
S=1+s_1M_{00}^{-1}+s_2M_{00}^{-2}+\ldots ,
$$
where 
$$
a=e^{c_1\tilde{x}_1+c_2}c_0
$$ 
(if such exponent exists, otherwise $a=c_0$), $0\ne c_0\in k$, $c_1\in k$, $c_2\in \ker\partial_1$ and $\tilde{x}_1\in A[[L_{00}^{-1}]]$ is an element satisfying 
$$
[L_{00},\tilde{x}_1]=0,\mbox{\quad} [\partial_2 +\partial_2(\log V),\tilde{x}_1]=0,
$$
$V\in 1+E_-$ is an operator such that $VL_{00}V^{-1}\in A[\partial_1]$, 
$$
s_n=\sum_{k=0}^n\tilde{x}_1^k\tilde{C}_k, \mbox{\quad} \tilde{C}_k=\sum_j a_jL_{00}^j\in \ker\partial_1((L_{00}^{-1})),
$$
the coefficients $a_j$ are defined recursively by formulas (\ref{adm7}), (\ref{adm5}) below.

The operator $T_0$ is the operator defined separately in each of the following two cases:

1) If $(M_{00}\partial_2^{-1})_+\in k((L_{00}^{-1}))$, then 
$T_0^{-1}=\exp((\tilde{x}_1k_1+l_1)L_{00}^{-1}+\ldots )$, where $k_i\in k$ satisfy the condition (\ref{adm12}) and $l_i\in \ker\partial_1$ are defined by condition (\ref{adm10}) below.

2) If $(M_{00}\partial_2^{-1})_+\not\in k((L_{00}^{-1}))$, then
$T_0^{-1}=\exp(l_1L_{00}^{-1}+l_2L_{00}^{-2}+\ldots )$, where  $l_i\in \ker\partial_1$ are defined by condition (\ref{adm13}) below.

\end{lemma}

{\bf Proof.} Every $(L_{00},M_{00})$-admissible operator $T\in E$ can be written as a product $T=S^{-1}T_0a$, where $S\in 1+E_{-}$, $T_0\in 1+A((L_{00}^{-1}))$, $a\in A$. 

First of all, let's consider the action of the operators $S^{-1}, T_0a$ on  $(L_{00},M_{00})$. For any operators $L,M$ as in theorem \ref{Parshin} we have 
\begin{equation}
\label{adm1}
(T_0aL_{+}a^{-1}T_0^{-1})_-=0, \mbox{\quad} (S^{-1}L_{+}S)_+=L_{+}
\end{equation}
by formulas (\ref{formulka}) and (\ref{formulka1}); and 
\begin{equation}
\label{adm15}
(T_0aM_{+}a^{-1}T_0^{-1})_-=0, \mbox{\quad} (S^{-1}M_{+}S)_+=M_+-x, \mbox{\quad} x\in A((L_{00}^{-1}))
\end{equation}
by formulas (\ref{formulka0}) and (\ref{formulka2}). Therefore we obtain two necessary conditions on the operator $T_0a$:
$$
T_0aL_{00}a^{-1}T_0^{-1}\in k((L_{00}^{-1})),
$$ 
$$
T_0av_{-1}a^{-1}T_0^{-1}v_{-1}^{-1}\in k((L_{00}^{-1})).
$$
Since $v_{-1}$ commute with $L_{00}$ and can be represented as a series in $L_{00}^{-1}$ with coefficients from $\ker \partial_1$ by lemma \ref{nichts}, the necessary conditions above imply that either coefficients of $v_{-1}$ belong to $k$ or $T_0a$ commute with $L_{00}$. So, we consider two cases:

{\it Case 1}. The coefficients of $v_{-1}$ belong to $k$.

{\it Case 2}. There are coefficients of $v_{-1}$ that do not belong to $k$. 

{\it Case 1}. Assume that the equation $\partial_1(x)=cx$ with $c\in A$ has a solution in the ring $A'\supset A$. Formally we will not need this assumption, as we will see later, but it makes the proof more convenient and compact. 

1) In this case there exists an invertible zeroth order operator $V\in E\otimes_A A'$ such that $VL_{00}V^{-1}=\partial_1$. Note that $V$ can be found in the form $V=V_0x$, where $V_0\in 1+E_-$ and $x\in A'$. Thus, $V$ preserve the ring $E$, $VEV^{-1}\subset E$.  
Obviously, the operator $T$ is $(L_{00}, M_{00})$-admissible if and only if $VTV^{-1}$ is $(VL_{00}V^{-1}, VM_{00}V^{-1})$-admissible. Replacing now the operators $L_{00}, M_{00}$ by $VL_{00}V^{-1}, VM_{00}V^{-1}$, we can assume $L_{00}=\partial_1$ and  $v_0$ is a polynomial in $\partial_1$ with coefficients from $\ker\partial_1(A)$ (see theorem \ref{Parshin}, iii)). 

Because of  (\ref{adm1}) we must have 
$$
\partial_1S=S(\partial_1 +k_1M_{00}^{-1}+k_2M_{00}^{-2}+\ldots ),
$$
where $k_i\in k((\partial_1^{-1}))$, where from 
\begin{equation}
\label{adm3}
\partial_1(s_i)=\sum_{q=0}^{i-1}s_qk_{i-q},
\end{equation}
where $S=1+s_1M_{00}^{-1}+s_2M_{00}^{-2}+\ldots$. Let's fix some element $x_1\in A$ with properties $\partial_1(x_1)=1$, $\partial_2(x_1)=0$. Such an element exists by the properties of the ring $A$, and it is defined up to a constant. 
So, from (\ref{adm3}) we obtain that  $s_i$ is a polynomial in $x_1$ with coefficients from $\ker\partial_1((\partial_1^{-1}))$, 
\begin{equation}
\label{adm2}
s_i=\sum_{q=0}^{i}x_1^qC_q,
\end{equation}
where 
$$
C_i=\sum_ja_{ji}\partial_1^j\in \ker\partial_1((\partial_1^{-1})).
$$ 
By (\ref{formulka0}) and (\ref{formulka2}) we have 
\begin{equation}
\label{adm25}
S^{-1}M_{00}S=M_{00}+[v_{-1}, s_1]v_{-1}^{-1}+(<0),
\end{equation}
and $[v_{-1}, s_1]$ does not depend on the choice of $s_1$. Since $v_{-1}$ is a series with constant coefficients and $s_1$ is a linear function in $x_1$, we obtain $[v_{-1}, s_1]\in k((\partial_1^{-1}))$. 

Therefore, by (\ref{adm25}) and (\ref{adm15}), the operators $S^{-1}, T_0a$ must be  $(L_{00},M_{00})$-admissible, and we can describe the conditions for these operators separately. 

2) For the operator $S$ we have the following condition:
$$
M_{00}S=SM_{00}+\sum_{q=1}^{\infty}[M_{00},s_q]M_{00}^{-q}=
$$
\begin{equation}
\label{adm4}
S(M_{00}+l_0+l_1M_{00}^{-1}+\ldots )=SM_{00}+l_0+\sum_{q=1}^{\infty}\sum_{m=0}^{q}s_ml_{q-m}M_{00}^{-q},
\end{equation}
where $l_i\in k((\partial_1^{-1}))$. We have by formula (\ref{formulka2})
\begin{equation}
\label{adm45}
[M_{00},s_q]=[v_{-1}, s_q]v_{-1}^{-1}M_{00}-[v_{-1},s_q]v_{-1}^{-1}v_0+v_{-1}\partial_2(s_q)+[v_0,s_q],
\end{equation}
where from we get 
\begin{equation}
\label{adm5}
\partial_2(s_q)=v_{-1}^{-1}(\sum_{m=0}^{q}s_ml_{q-m} - [v_0,s_q] +[v_{-1},s_q]v_{-1}^{-1}v_0 - [v_{-1}, s_{q+1}]v_{-1}^{-1})
\end{equation}
for all $q\ge 1$, and $l_0=[v_{-1}, s_{1}]v_{-1}^{-1}=[v_{-1}, x_{1}]v_{-1}^{-1}k_1$. Note that we have
\begin{equation}
\label{adm55}
[v_0,s_q]=\sum_{m=0}^{q-1}x_1^mB_m, \mbox{\quad} [v_{-1},s_q]v_{-1}^{-1}v_0=\sum_{m=0}^{q-1}x_1^mB'_m,
\end{equation}
where $B_{m}, B'_m\in \ker\partial_1((\partial_1^{-1}))$ depend on $C_{m+1},\ldots C_q$ and do not depend on $C_j$, $j\le m$. Moreover, we have 
\begin{equation}
\label{adm6}
[v_{-1}, s_{q+1}]v_{-1}^{-1}=(\sum_{m=0}^{q}s_mk_{q+1-m})
[v_{-1}, x_1]v_{-1}^{-1}+\sum_{m=0}^{q-1}x_1^mH_m,
\end{equation}
where $H_m\in \ker\partial_1((\partial_1^{-1}))$ depend on $C_{m+1},\ldots C_q$ and do not depend on $C_j$, $j\le m$ (here $C_j$ are the coefficients of the polynomial $s_q$). So, we obtain 
$$
s_ql_0-[v_{-1}, s_{q+1}]v_{-1}^{-1}=(\sum_{m=0}^{q-1}s_mk_{q+1-m})
[v_{-1}, x_1]v_{-1}^{-1}+\sum_{m=0}^{q-1}x_1^mH_m,
$$
where from, together with (\ref{adm55}), the formula (\ref{adm5}) can be rewritten as
\begin{equation}
\label{adm7}
\sum_{m=0}^{q-1}x_1^m(\sum_j\partial_2(a_{jm})\partial_1^j)=\sum_{m=0}^{q-1}x_1^mW_m,
\end{equation}
where $W_m\in \ker\partial_1((\partial_1^{-1}))$ depend on $C_{m+1},\ldots C_q$ and do not depend on $C_j$, $j\le m$. 

So, the equation (\ref{adm7}) is solvable for arbitrary $q\ge 0$, and the set of all $(L_{00},M_{00})$-admissible operators from the group $1+E_-$ can be described as a set of all operators $S$ with coefficients $s_q$ of the form (\ref{adm2}) defined recursively by formulas  (\ref{adm3}), (\ref{adm7}), (\ref{adm5}) for arbitrary parameters $k_i,l_i\in k((L_{00}^{-1}))$, $i\ge 1$.     

Note that formula (\ref{adm7}) actually defines the {\it coefficients} of the series $C_m$, which belong to $\ker\partial_1(A)$. So, conjugating  operators $S$ with the operator $V$, we obtain the same description for admissible operators in general case just setting $\tilde{x}_1:=V^{-1}x_1V\in A((\partial_1))$ instead of $x_1$. Actually $[L_{00},\tilde{x}_1]=0$ and $[\partial_2 +\partial_2\log V,\tilde{x}_1]=0$.

3) Now for the operator $S_1^{-1}:=T_0a$ we have the following conditions:
\begin{equation}
\label{adm8}
S_1^{-1}\partial_1S_1=\partial_1+\partial_1(\log S_1)=\partial_1+k_0+\ldots \in k((\partial_1^{-1})),
\end{equation}
where from $\log S_1=(x_1k_0+l_0)+(x_1k_1+l_1)\partial_1^{-1}+\ldots $, where $l_i\in \ker\partial_1$, and
\begin{equation}
\label{adm9}
S_1^{-1}M_{00}S_1=(S_1^{-1}v_{-1}S_1v_{-1}^{-1})M_{00}-(S_1^{-1}v_{-1}S_1v_{-1}^{-1})v_0+S_1^{-1}v_0S_1+(S_1^{-1}v_{-1}S_1)\partial_2(\log S_1),
\end{equation} 
where from 
$$
\partial_2(\log S_1)=
$$
\begin{equation}
\label{adm10}
\sum_{i=0}^{\infty}\partial_2(l_i)\partial_1^{-i}\in k((\partial_1^{-1}))+(S_1^{-1}v_{-1}^{-1}S_1)((S_1^{-1}v_{-1}S_1v_{-1}^{-1})v_0-S_1^{-1}v_0S_1)\in \ker\partial_1((\partial_1^{-1})).
\end{equation}
Clearly, this equation is solvable only if 
\begin{equation}
\label{adm11}
v_{-1}^{-1}v_0-S_1v_{-1}^{-1}v_0S_1^{-1}\in \ker\partial_1[[\partial_1^{-1}]] \mbox{\quad}\mod k((\partial_1^{-1})).
\end{equation}

From two conditions (\ref{adm10}), (\ref{adm8}) we obtain that the coefficient $a$ of all possible operators $T_0a$ must be of the form $e^{c_1x_1+c_2}c_0$ (if such an exponent exists), where $c_0\in k^{\times}$, $c_1\in k$ and $c_2\in \ker\partial_1$. Then the operator $T_0^{-1}$ must be of the form 
$T_0^{-1}=\exp((x_1k_1+l_1)\partial_1^{-1}+\ldots )$, where $k_i\in k$ satisfy the condition 
\begin{equation}
\label{adm12}
v_{-1}^{-1}v_0-T_0^{-1}v_{-1}^{-1}v_0T_0\in (\ker\partial_1[[\partial_1^{-1}]])\partial \mbox{\quad}\mod k((\partial_1^{-1})),
\end{equation}
and $l_i\in \ker\partial_1$ are defined by condition (\ref{adm10}). Obviously, the last condition is always compatible with any choice of parameters $k_i$. Note also that (\ref{adm12}) don't depend on $l_i$, so our description is well defined.

Replacing $x_1$ with $\tilde{x}_1$ and $\partial_1$ with $L_{00}$ in formulas (\ref{adm12}), (\ref{adm10}), we obtain, as in the case of operators $S$, the description of operators $T_0$ in general case.

{\it Case 2.} This case is very similar to the case 1. The difference is that the operators $S, T_0a$ can be not $(L_{00},M_{00})$-admissible. Repeating the proof of the case 1, we obtain the same conditions for the operator $S$, except the condition $l_0\in k((\partial_1^{-1}))$. Now $l_0\in \ker\partial_1((\partial_1^{-1}))$. This gives us a new condition for the operator $T_0a$. Namely, instead of condition (\ref{adm11}), which holds automatically because $T_0a$ commutes with $\partial_1$, we obtain 
\begin{equation}
\label{adm13}
\partial_2(\log S_1)=\sum_{i=0}^{\infty}\partial_2(l_i)\partial_1^i\in v_{-1}^{-1}(k((\partial_1^{-1}))-[v_{-1},s_1]v_{-1}^{-1}).
\end{equation}

Replacing $x_1$ with $\tilde{x}_1$ and $\partial_1$ with $L_{00}$  we obtain  the description of operators $T_0$ in general case.

The lemma is proved.\\
$\Box$

\subsection{Solvability of the modified SW-systems}

{\it In this subsection $A$ need not be commutative}. 

To find a solution of the Cauchy problem for the modified Sato-Wilson systems we can follow the way described in \cite{Mu}. Of course, for the modified systems the proofs will be more complicated.

The following theorem is a generalization of theorem 4.1. in \cite{Mu}. 

\begin{theo}
\label{4.1}
Let $\Omega_{\alpha}=\sum_{i\in\sdz,j\in\sdz_+,i\le\alpha j}P_{ij}dt_{ij}$ be a $D_R$-valued 1-form satisfying 

a) there is a positive real number $c>0$ such that 
$$
\ord_{\partial_2}P_{ij}\le \frac{j}{c} \mbox{\quad for all $i,j$}.
$$

b) $P_{ij}=P_{ij}|_{t_{m,n}=0, m>\alpha n}$. 

c) $\Omega_{\alpha}$ is integrable, i.e. $d_{\alpha}\Omega_{\alpha}= \Omega_{\alpha}\wedge \Omega_{\alpha}$. 

Then for every given operator $Y(0)\in  E_{+}$, there may exist only one  solution $Y\in \widehat{D}_R$ of the linear total differential equation 
$$
d_{\alpha}Y=\Omega_{\alpha}Y
$$ 
having $Y(0)$ as its initial value; $Y|_{t=0}=Y(0)$ and $Y|_{ t_{m,n}=0, m>\alpha n}=Y$. 
\end{theo}

{\bf Remark.} This version of theorem is weaker than in \cite{Mu} or in \cite{Zh}, Th. 1. Nevertheless, for our aim this version is sufficient. The existence of a solution of a special system from this theorem will follow from a generalized Birkhoff decomposition, which we will prove below. 

\smallskip

{\bf Proof.} Let's fix some bijection between the set of all pairs $(i,j)$ with $i\in\dz$, $j\in\dz_+$, $i\le \alpha j$ and natural numbers, 
$\zeta : \{(i,j)\}\longrightarrow \dn$, such that $\zeta (i,j)=k$, where $k\ge j$. Such a bijection can be constructed, for example, by counting integral points (lying below the graphic of the function $\alpha$) lying inside  polygons formed by parallel lines that go through the points $(i,0)$, $(i-1,0)$, $i\le \alpha (0)$, 
and are parallel to the line going through the points $(\alpha (0)-1,0)$ and $(\alpha (1), 1)$, the vertical lines $(.,k)$ and lines going through the points $(\alpha (k),k)$, $(\alpha (k+1),k+1)$.

Now, because of the condition b), the differential equation from the theorem can be rewritten as  
\begin{equation}
\label{ttt}
dY= (\sum_{n\ge 1}^{\infty}  P_{\zeta^{-1}(n)}dt_{\zeta^{-1}(n)})Y, \mbox{\quad} d=\sum_{n=1}^{\infty}dt_{\zeta^{-1}(n)}\frac{\partial}{\partial t_{\zeta^{-1}(n)}}
\end{equation}
where $P_{\zeta^{-1}(n)}$ are considered as the elements with variables $t_{\zeta^{-1}(j)}$ instead of $t_{m,n}$. 

Let's define the ring ${\cal R}=A((\partial_1^{-1}))[[t_{\zeta^{-1}(1)}, t_{\zeta^{-1}(2)}, \ldots ]]$ as a projective limit ring with a pseudo-valuation $v'_2:{\cal R}\backslash\{0\}\rightarrow \dn$ defined by $v'_2(t_{\zeta^{-1}(n)})=n$. Let $\cal D$, $\cal E$, $\widehat{\cal D}$, $\widehat{\cal E}$ be rings defined with respect to ${\cal R}$ on pages 9,10 in \cite{Mu}. 

Because of the valuation growth conditions on coefficients of elements from the ring $\widehat{E}_{\bar R}$ and because of special choice of the bijection, the elements $P_{\zeta^{-1}(n)}$  can be represented as series from the ring $\cal D$. Analogously, every element $Y\in \widehat{D}_R$ with the properties as in the theorem can be represented as an element from $\widehat{\cal D}$. Obviously, such a representation is uniquely defined, that is two different elements from $\widehat{D}_R$ give two different elements from the ring $\widehat{\cal D}$. Of course, the representation and the equation (\ref{ttt}) depend on the choice of bijection. 

By the property of the bijection
$$
\ord_{\partial_2}P_{\zeta^{-1}(n)}=\ord_{\partial_2}P_{i,j}\le \frac{j}{c}\le \frac{n}{c}
$$
for all $n$. So, the equation (\ref{ttt}) is the differential equation from theorem 4.1 in \cite{Mu}, and therefore have a unique solution for any initial value $Y(0)\in E_+$. As we have seen, any solution of the equation from the theorem gives a uniquely defined 

Now, if there were two different solutions of the differential equation from the theorem, the equation (\ref{ttt}) would have two different solutions  with the same initial value, a  contradiction. \\
$\Box$

{\bf Remark.} Since an element from $\widehat{\cal D}$ can not be in general represented as an element of $\widehat{D}_R$, the existence of a solution of the equation (\ref{ttt}) in $\widehat{\cal D}$ do not imply the existence of a solution of the differential equation from the theorem. 

\smallskip

Now we can formulate a generalized version of the Birkhoff decomposition theorem. 

\begin{theo}
\label{birkhoff}
For any element $U\in {\widehat{E}_R}^{\times}$ there exists a unique factorization 
$$
U=S^{-1}Y,
$$
where $S\in \widehat{\Nu}_R$ and $Y\in {\widehat{D}_R}^{\times}$.
\end{theo}

{\bf Proof.} The proof of this theorem is analogous to the proof of Theorem 3.2. in \cite{Mu} with certain modifications. 

The uniqueness is trivial. Indeed, if $S_1^{-1}Y_1=S_2^{-1}Y_2$, then $S_1S_2^{-1}=Y_1Y_2^{-1}\in \widehat{\Nu}_R\cap \widehat{D}_R^{\times}=\{1\}$. Hence $S_1=S_2$ and $Y_1=Y_2$. To show the existence of the decomposition, we have to solve the equation $SU\in \widehat{D}_R$ for an unknown $S\in \widehat{\Nu}_R$. 

So let $U=\sum_{\beta\in \sdz}u_{\beta}\partial_2^{\beta}$ and 
$S=1+\sum_{\gamma =1}^{\infty}s_{\gamma}\partial_2^{-\gamma}$, where 
$u_{\beta},s_{\gamma}\in \widehat{E}_{\bar{R}}$. 
Then 
\begin{multline*}
SU=(1+\sum_{\gamma =1}^{\infty}s_{\gamma}\partial_2^{-\gamma})(\sum_{\beta\in\sdz}u_{\beta}\partial_2^{\beta}) \\
=\sum_{\beta\in\sdz}u_{\beta}\partial_2^{\beta}+\sum_{\gamma =1}^{\infty}\sum_{\beta\in\sdz}\sum_{i=0}^{\infty}C_{-\gamma}^is_{\gamma}u_{\beta}^{(i)}\partial_2^{-\gamma +\beta -i}\\
=\sum_{\beta\in\sdz}u_{\beta}\partial_2^{\beta}+\sum_{\delta\in\sdz}(\sum_{\gamma =1}^{\infty}\sum_{i=0}^{\infty}C_{-\gamma}^is_{\gamma}u_{\delta +\gamma +i}^{(i)})\partial_2^{\delta},
\end{multline*}
where $u^{(i)}=\partial_2^i(u)$. 

Therefore the equation we have to solve is a system of algebraic equations
\begin{equation}
\label{3.80}
u_{-\beta}+\sum_{\gamma =1}^{\infty}\sum_{i=0}^{\infty}C_{-\gamma}^is_{\gamma}u_{\gamma -\beta +i}^{(i)}=0 \mbox{\quad for $\beta =1,2,3,\ldots$}
\end{equation} 
Define
$$
\bfu =(u_{-1},u_{-2},u_{-3},\ldots ), \mbox{\quad } \bfs =(s_1,s_2,s_3, \ldots )
$$
and 
\begin{equation}
\label{Matr0}
M=\left [
\sum_{i=0}^{\infty}C_{-\gamma}^iu_{\gamma -\beta +i}^{(i)}\right ]{}_{\gamma ,\beta =1,2,3, \ldots},
\end{equation}
where $M$ is a square matrix of infinite size with coefficients in $\widehat{E}_{\bar{R}}$. 
Equation $(\ref{3.8})$ now reads 
\begin{equation}
\label{3.90}
\bfs M=-\bfu
\end{equation}
Therefore the solution $\bfs$ is given by $\bfs =-\bfu M^{-1}$. The idea is to define $M^{-1}$ by the  Neumann series $\sum_{n=0}^{\infty}(1-M)^n$, and use a similar technique developed in the proof of lemma \ref{3.1} to establish well-definedness of $\sum_{n=0}^{\infty}(1-M)^n$ and $\bfu M^{-1}$. Since $\bfs$ determines all the coefficients of $S$, well-definedness of $\bfu\sum_{n=0}^{\infty}(1-M)^n$ implies the existence of $S$ such that $SU\in   \widehat{D}_R$. 

Let $Q=1-M=[a_{\mu\nu}]_{\mu ,\nu=1,2,3,\ldots}$ and $Q^n=[a_{n,\mu\nu}]_{\mu ,\nu =1,2,3,\ldots }$. Since $a_{\mu\nu}=\delta_{\mu\nu}-\sum_{i=0}^{\infty}C_{-\mu}^iu_{\mu -\nu +i}^{(i)}$, we have 
$$
 (a_{\mu\mu})|_{t=0}=1-(\sum_{i=0}^{\infty}C_{-\mu}^iu_i^{(i)})|_{t=0}=0,
$$
because of definition of $\widehat{E}_R^{\times}$. Similarly, if $\mu >\nu$, then 
$$
\pi (a_{\mu\nu})=\pi (-\sum_{i=0}^{\infty}C_{-\mu}^iu_{\mu -\nu +i}^{(i)})=0.
$$
Because of the growth order condition for $u_{\nu}$'s, we can find a positive real number $J$ such that 
\begin{equation}
\label{3.1000}
v_2(a_{\mu\nu})\ge J(\mu -\nu ) \mbox{\quad for all $\mu -\nu\ge 0$}.
\end{equation}
as before.
 
{\it Claim.} i) For every $n\ge 1$ we have $v_2(a_{n,\mu \nu})\ge J(\mu -\nu )$ if $\mu -\nu \ge 0$. 

ii) The function $F(M,N,a_{n,\mu \nu})$ satisfy the following property: for any given $M,N,\mu ,\nu$ there exists a natural number $T(M,N,\mu ,\nu )$ such that $F(M,N,a_{n,\mu \nu})=0$ for all $n>T(M,N,\mu ,\nu )$.

Here we take the function $F$ as in the proof of lemma \ref{3.1}.   

{\it Proof of the claim.} 1) Let's prove the first assertion. If $n=1$, then $a_{1,\mu \nu}=a_{\mu \nu}$ and i) follows from (\ref{3.1000}). Assume that i) holds for some $n\ge 1$. Since $Q^{n+1}=Q^n\cdot Q$,
\begin{equation}
\label{bir1}
a_{n+1,\mu \nu}=\sum_{l=1}^{\infty}a_{n,\mu l}a_{l\nu}= \sum_{l=1}^{\nu -1}a_{n,\mu l}a_{l\nu}+ \sum_{l=\nu}^{\mu }a_{n,\mu l}a_{l\nu} +\sum_{l=\mu +1}^{\infty}a_{n,\mu l}a_{l\nu}.
\end{equation}
We assume that $\mu -\nu\ge 0$. In the first term of (\ref{bir1}), since $\nu >l$, we have $\mu -l>\mu -\nu\ge 0$, and hence
$$
v_2(a_{n,\mu l}a_{l\nu})\ge v_2(a_{n,\mu l})\ge J(\mu -l)>J(\mu -\nu ).
$$
Therefore 
$$
v_2(\sum_{l=1}^{\nu -1}a_{n,\mu l}a_{l\nu})\ge J(\mu -\nu ).
$$
In the second term of (\ref{bir1}), since $\nu\le l\le \mu$, we have 
$$
v_2(a_{n,\mu l}a_{l\nu})=v_2(a_{n,\mu l})+v_2(a_{l\nu})\ge J(\mu -l)+J(l-\nu )=J(\mu -\nu).
$$
Therefore, 
$$
v_2(\sum_{l=\nu}^{\mu }a_{n,\mu l}a_{l\nu})\ge J(\mu -\nu ).
$$
Finally, in the third item of (\ref{bir1}), $l>\mu $ implies $l-\nu >\mu -\nu$. Therefore, 
$$
v_2(\sum_{l=\mu +1}^{\infty}a_{n,\mu l}a_{l\nu})\ge v_2(a_{l\nu})\ge J(\mu -\nu ).
$$

2) Now let's prove the second assertion. To prove it we use an induction on $N$ and $|\mu -\nu|$, like in the proof of lemma \ref{3.1}. We will use here the function $q(N,a)$ and various properties used there. 

It suffice to prove the assertion only for $M<0$. 
Let's prove the first step of our double induction. Let $N=0$ and $|\mu -\nu|=0$. Then we have $T(M,N,\mu ,\nu )=|M|$ and $q(N,a_{n,\mu \nu})\le -n$. Indeed, by (\ref{bir1}) 
$$
F(M,0,a_{n+1,\mu \mu})=F(M,0,a_{n,\mu \mu}a_{\mu \mu})=\ldots =F(M,0,a_{\mu \mu}^{n+1}),
$$ 
because $v_2(a_{j,\mu \nu})\ge J(\mu -\nu )>0$ for all $j,(\mu -\nu )\ge 1$ by the first assertion of the claim, so all the summands in (\ref{bir1}) except $a_{n,\mu \mu}a_{\mu \mu}$ have valuation greater than zero and therefore they don't change the value of the function $F(M,0,.)$. Obviously, the same is true for the function $q(0,.)$. Since $q(0, a_{\mu \mu})\le -1$, we have by (\ref{tex2}) $a_{\mu \mu}^n\le -n$, where from $F(M,0,a_{\mu \mu}^{n+1})=0$ for all $n>|M|$ and $q(0,a_{n+1,\mu \mu})\le -n-1$. 

3) Now let $N=0$ and $|\mu -\nu |>0$. Since $v_2(a_{j,\mu \nu})>0$ for all $j,(\mu -\nu )\ge 1$, we have $F(M,0, a_{n,\mu \nu})=0$, $q(0,a_{n,\mu \nu})=-\infty$ for all $n$ and $(\mu -\nu )\ge 1$. So, we can assume $(\mu -\nu )<0$. 

By the induction hypothesis $F(M,0,a_{n,\mu' \nu'})=0$ for all $n>T(M,0,\mu' ,\nu' )$ if $|\mu' -\nu' |<|k|$. Denote by $N_0$ the maximum of all $T(M,0,\mu l ), T(M,0,l \nu )$ with $\mu <l<\nu$ (with some fixed $\mu ,\nu$, $|\mu -\nu |=|k|$). Now consider the elements $a_{n+N_0,\mu \nu}$ with $n>N_0$. 
Since $Q^{n+N_0}=Q^nQ^{N_0}$, we have by the same reason as above
\begin{equation}
\label{bir6}
F(M,0,a_{n+N_0,\mu \nu})=
F(M,0,\sum_{l=\mu}^{\nu}a_{n,\mu l}a_{N_0,l \nu})\le \max_{\mu \le l\le \nu }\{F(M,0,a_{n,\mu l}a_{N_0,l \nu})\},
\end{equation}
where the last inequality follows from the property ii) of the function $F$, see step a) of theorem \ref{Parshin}.

4) So, it suffice to prove the assertion for each function $F(M,0,a_{n,\mu l}a_{N_0,l \nu}))$, where $ \mu \le l\le \nu$, and then define the number $T(M,0,\mu ,\nu )$ as a maximum of corresponding numbers for each $l$. 

For $ \mu < l< \nu$  we have $F(M,0,a_{n,\mu l})=0$ and $F(M,0,a_{N_0,l \nu})=0$. Therefore, by (\ref{tex4}) we obtain $F(M,0,a_{n,\mu l}a_{N_0,l \nu})=0$ for all $n>N_0$. 

Now consider the case $l=\mu $.  
By the property (\ref{tex2}) we have 
$$
q(0,a_{n,\mu \mu}a_{N_0,\mu \nu })\le q(0,a_{n,\mu \mu})+q(0,a_{N_0,\mu \nu})\le -n+q(0,a_{N_0,\mu \nu}).
$$
Therefore, for all $n>-M+q(0,a_{N_0,\mu \nu})$ we have $F(M,0,a_{n,\mu \mu}a_{N_0,\mu \nu })=0$. 

5) Combining all together, we get for all $n>\max\{N_0,-M+q(0,a_{N_0,\mu \nu})\}:=\tilde{T}$ (see (\ref{bir6})):
\begin{equation}
\label{bir7}
F(M,0,a_{n+N_0,\mu \nu})\le F(M,0,a_{n,\mu \nu}a_{N_0,\nu \nu}).
\end{equation}

Now put 
$$O=(\max_{\tilde{T}\le j\le \tilde{T}+N_0}\{F(M,0,a_{j,\mu \nu})\})$$ 
and put $T(M,0,\mu ,\nu )=\tilde{T}+ON_0$. We claim, that for all $n>T(M,0,\mu ,\nu )$ $F(M,0,a_{n,\mu \nu})=0$. 

Assume the converse. Let $n>T(M,0,\mu ,\nu )$ be the number such that $F(M,0,a_{n,\mu \nu})>0$. Then by (\ref{bir7}) and (\ref{tex2}) we have 
$$
0<F(M,0,a_{n,\mu \nu})=q(0,a_{n,\mu \nu})-M\le F(M,0,a_{n-N_0,\mu \nu}a_{N_0,\nu \nu})=q(0,a_{n-N_0,\mu \nu}a_{N_0,\nu \nu})-M
$$
$$
\le q(0,a_{n-N_0,\mu \nu})+q(0,a_{N_0,\nu \nu})-M\le q(0,a_{n-N_0,\mu \nu})-M -N_0=F(M,0,a_{n-N_0,\mu \nu})-N_0\le \ldots 
$$
$$
\le F(M,0,a_{n-\tilde{O}N_0,\mu \nu})-\tilde{O}N_0,
$$
where $\tilde{T}\le n-\tilde{O}N_0\le \tilde{T}+N_0$ and therefore $\tilde{O}>O$. So, 
$$
F(M,0,a_{n-\tilde{O}N_0,\mu \nu})-\tilde{O}N_0\le O-\tilde{O}N_0\le 0,
$$
a contradiction. 

6) Now assume $N$ is an arbitrary positive number. Let $k_0:=[N/J]$ be the
integral part of the number $N/J$. For all $\mu -\nu >k_0$ we have $v_2(a_{n,\mu \nu})\ge J(\mu -\nu )>N$, so the assertion is trivial for all such $(\mu -\nu )$. 

To prove the assertion we will use an inverse induction on $(\mu -\nu )\le k_0$. 
By the induction hypothesis on $N$ there exists a natural number $N_0$ such that $F(M,N-1,a_{n,\mu l})=0, F(M,N-1,a_{n,l \nu})=0$ for all $n\ge N_0$ and $\nu\le l\le \mu$ (with fixed $M,\mu ,\nu$). We then have for all $n>N_0$ and $\mu -\nu =k_0$ 
$$
F(M,N,a_{n+N_0,\mu \nu })=
F(M,N,\sum_{l=\nu}^{\mu}a_{n,\mu l}a_{N_0,l \nu})\le 
$$
$$
\max_{\nu \le l\le \mu}\{F(M,N,a_{n,\mu l}a_{N_0,l \nu})\}\le 
\max\{F(M,N,a_{n,\mu \mu}a_{N_0,\mu \nu}), F(M,N,a_{n,\mu \nu}a_{N_0,\nu \nu})\},
$$
where the last inequality follows from (\ref{tex5}). For all $n>-M+q(N,a_{N_0,\mu \nu})$ we have, as above, $F(M,N,a_{n,\mu \mu}a_{N_0,\mu \nu})=0$. So, for all sufficiently large $n$ 
$$
F(M,N,a_{n+N_0,\mu \nu })\le F(M,N,a_{n,\mu \nu}a_{N_0,\nu \nu}),
$$
and we can repeat the arguments of step 7) to get the proof of the assertion in the case $\mu -\nu =k_0$. 

7) Let's prove the assertion for arbitrary $\mu -\nu <k_0$. By the induction hypothesis on $N, k=(\mu -\nu )$ there exists a natural number $N_0$ such that $F(M,N-1,a_{n,\mu j})=0, F(M,N-1,a_{n,j \nu})=0$ for all $n\ge N_0$ and $\nu +k_0\ge j\ge \mu-k_0$, $F(M,N,a_{n,\mu j})=0$ for all $n\ge N_0$ and $\nu > j\ge \mu -k_0$, and $F(M,N,a_{n,j \nu})=0$ for all $n\ge N_0$ and $\nu +k_0\ge j> \mu $. We then have for all $n>N_0$
$$
F(M,N,a_{n+N_0,\mu \nu})=
F(M,N,\sum_{l=\mu-k_0}^{\nu +k_0}a_{n,\mu l}a_{N_0,l \nu})\le 
$$
$$
\max_{\mu-k_0\le l\le \nu +k_0}\{F(M,N,a_{n,\mu l}a_{N_0,l \nu})\}\le 
\max\{F(M,N,a_{n,\mu \mu}a_{N_0,\mu \nu})), F(M,N,a_{n,\mu \nu}a_{N_0,\nu \nu})\},
$$
where the last inequality follows from (\ref{tex4}), (\ref{tex5}), (\ref{tex51}), (\ref{tex52}). Indeed, if $l\ne \mu ,\nu$ we have either $\mu -l<k$ and therefore $l-\nu >0$ or $l-\nu <k$ and therefore $\mu -l>0$ or $\mu -l>k$, $l-\nu >k$. In the first case we apply (\ref{tex5}) or (\ref{tex52}), in the second case we apply (\ref{tex5}) or (\ref{tex51}), and in the third case we apply (\ref{tex4}). 

Repeating the arguments of the end of step 8), we obtain the assertion also in this case, hence in general. \\
The claim is proved.

\smallskip

By this claim we can conclude that $\sum_{n=0}^{\infty}a_{n,\mu \nu}\in \widehat{E}_{\bar R}$ is well defined for all $\mu ,\nu =1,2,3,\ldots $. Therefore, 
$$
M^{-1}=\sum_{n=0}^{\infty}(1-M)^n=
\left [
\sum_{n=0}^{\infty}a_{n,\mu \nu}
\right ]_{\mu ,\nu}
$$
is well defined. Let $M^{-1}=[b_{\mu \nu}]_{\mu ,\nu =1,2,\ldots }$, namely $b_{\mu \nu}=\sum_{n=0}^{\infty}a_{n,\mu \nu}$. If $\mu >\nu$, then $v_2(b_{\mu \nu})=v_2(\sum_{n=0}^{\infty}a_{n,\mu \nu})\ge J(\mu -\nu )$. Therefore
$$
s_{\nu}=-\sum_{\mu =1}^{\infty}u_{-\mu}b_{\mu \nu}=-\sum_{\mu =1}^{\nu}u_{-\mu}b_{\mu \nu}-\sum_{\mu =\nu +1}^{\infty}u_{-\mu}b_{\mu \nu}
$$
is a well-defined element in $\widehat{E}_{\bar R}$. Thus we have established the existence of $S\in \widehat{\Nu}_R$ such that $SU\in\widehat{D}_R$.

Finally, let $Y=SU$. Then $Y\in \widehat{D}_R^{\times}$ and $U=S^{-1}Y$. This completes the proof of the theorem.\\
$\Box$

Let us illustrate how we use the Birkhoff factorization to solve the systems $(SW)_{\alpha}$. To solve the system $(SW)_{\alpha}$ with the initial value $S(0)\in 1+E_{-}$ we take the explicit solution of the system
\begin{equation}
\label{1.29'}
d_{\alpha}U=\omega_{\alpha}^{N_0}U,
\end{equation}
given by 
\begin{equation}
\label{1.31}
U=\exp (\sum_{i\in \sdz ,j\in \sdz_+, i\le\alpha j}t_{i,j}L_{00}^iM_{00}^j)S(0)^{-1}\in \widehat{E}_R^{\times}
\end{equation}
and find, according to theorem \ref{birkhoff}, its unique decomposition $U=S^{-1}Y$. Define 
$$
Z_{\pm}^{\alpha}=\pm\sum_{i\le \alpha j}dt_{i,j}(SL_{00}^iM_{00}^jS^{-1})_{\pm}.
$$
We have 
$$
Z_{+}^{\alpha}-Z_{-}^{\alpha}=S\omega_{\alpha}S^{-1}=SdUU^{-1}S^{-1}=dYY^{-1}-dSS^{-1}
$$
Since $\widehat{D}_R\cap \widehat{E}_{R-}=\{0\}$, we obtain $Z_{+}^{\alpha}=dYY^{-1}$ and $Z_{-}^{\alpha}=dSS^{-1}$. So, $S$ gives a solution of $(SW)_{\alpha}$. Note that $U$ (so, $S$) does not depend on $t_{i,j}$, $i>\alpha j$. Since $Y|_{t=0}=1$, 
$$
S(0)^{-1}=U|_{t=0}=(S^{-1}|_{t=0})(Y|_{t=0})=(S|_{t=0})^{-1},
$$
namely, $S|_{t=0}=S(0)$. 

Now let's prove the uniqueness of a solution $S$. Let $S'\in\widehat{\Nu}_R$ be another solution of $(SW)_{\alpha}$ with the initial value $S(0)$ such that $S'$ does not depend on $t_{i,j}$, $i>\alpha j$. Then we also have $Z'^{\alpha}_{\pm}=\pm\sum_{i\le \alpha j}dt_{i,j}(S'L_{00}^iM_{00}^jS'^{-1})_{\pm}$ satisfying, by propositions \ref{SW}, \ref{svyaznost'} , $d_{\alpha}Z'^{\alpha}_{\pm}=Z'^{\alpha}_{\pm}\wedge Z'^{\alpha}_{\pm}$, and $Z'^{\alpha}_{\pm}$ do not depend on $t_{i,j}$, $i>\alpha j$. As we have seen  in the proof of theorem \ref{4.1}, the system  
$$
d_{\alpha}Y'=Z'^{\alpha}_{+}Y'
$$
can be represented, with help of some fixed bijection $\zeta$, as a system (\ref{ttt}), which has a unique solution in  the ring $\widehat{\cal D}^{\times}$. The original solution $S$ and the solution $S'$ of the system $(SW)_{\alpha}$ can be also uniquely represented as elements of $\widehat{\cal E}^{\times}$ with help of $\zeta$, as one can easily check. So, if we define $U'=S'^{-1}Y'\in \widehat{\cal E}^{\times}$, $U'$ must satisfy the system of linear partial differential equations:
\begin{equation}
\label{1.29}
dU'=\omega_{\alpha}^{N_0}U', \mbox{\quad} d=\sum_{n=1}^{\infty}dt_{\zeta^{-1}(n)}\frac{\partial}{\partial t_{\zeta^{-1}(n)}}.
\end{equation}
Indeed, 
\begin{multline*}
dU'=-S'^{-1}dS'S'^{-1}Y'+S'^{-1}dY'=S'^{-1}(dY'Y'^{-1}-dS'S'^{-1})S'S'^{-1}Y'\\
=S'^{-1}(Z'^{\alpha}_{+}-Z'^{\alpha}_{-})S'U'=S'^{-1}\sum_{i\le \alpha j}dt_{i,j}S'L_{00}^iM_{00}^j S'^{-1}S'U'=\omega_{\alpha}^{N_0}U'.
\end{multline*}

\begin{lemma}
\label{4.3}
If the equation (\ref{1.29}) has two solutions $U$ and $V\in \widehat{\cal E}$ with the same initial value $U|_{t=0}=V|_{t=0}$, then $U=V$.
\end{lemma}

{\bf Proof.} Let $W=U-V$. Note that $W|_{t=0}=U|_{t=0}-V|_{t=0}=0$. Since (\ref{1.29}) is linear, we have $dW=\omega_{\alpha}^{N_0}W$, namely, $\partial W/\partial t_{\zeta^{-1}(n)}=L_{00}^iM_{00}^jW$. Therefore, 
$$
\frac{\partial }{\partial t_{\zeta^{-1}(n_1)}}\ldots \frac{\partial }{\partial t_{\zeta^{-1}(n_k)}}W=L_{00}^{i_1+\ldots +i_k}M_{00}^{j_1+\ldots +j_k}W
$$ 
for any $n_1,\ldots n_k$,
where from, since $[M_{00},W]|_{t=0} =[M_{00},W|_{t=0}]$, we have 
$$
(\frac{\partial }{\partial t_{\zeta^{-1}(n_1)}}\ldots \frac{\partial }{\partial t_{\zeta^{-1}(n_k)}}W)(0)=
L_{00}^{i_1+\ldots +i_k}M_{00}^{j_1+\ldots +j_k}\cdot (W|_{t=0})=0.
$$
This means that $W$ does not depend on $t$. Since $W|_{t=0}=0$, we can conclude that $W=0$. \\
$\Box$

So, we obtain $S'^{-1}Y'=S^{-1}Y$ in the ring $\widehat{\cal E}^{\times}$. Since the Birkhoff decomposition is unique, we conclude $S'=S$ and $Y'=Y$. Therefore, $S'=S$ and $Y'=Y$ also in the ring $\widehat{E}_R^{\times}$. This completes the proof of the following theorem, which is an analog of the theorem 1.4 in \cite{Mu}:
\begin{theo}
\label{1.4}
For every initial value $S(0)\in 1+E_{-}$, there is a unique solution $S=S(t)\in \widehat{\Nu}_R$ of the system $(SW)_{\alpha}$ such that $S|_{t=0}=S(0)$. 
\end{theo}

$\Box$

\section{Existence of non-trivial solutions for the modified systems}

In this section we will look for an answer on the following question:
when the solution $S(t)$ of the system $(SW)_{\alpha}$ corresponds to a non-trivial solution of the system $(KP)_{\alpha}$ and belongs to the group $\Nu_R$. This question in general seems to be  very difficult, and we don't know the answer. So, we will consider here only the case of a linear function $\alpha :j\mapsto \alpha \cdot j$, $\alpha\in \dr$. 

As we have seen in the section 2.1, in the case 
$\alpha =\infty$ the original system $(KP)_{\alpha}$ has only trivial solutions. 
The theorem below shows that $S(t)\in \Nu_R$ for some initial values $S(0)$. 
The theorem does not cover all possible initial values, but nevertheless, the  examples described below will be important for applications. Other examples see in \cite{Zh}, section 4.

\begin{theo}
\label{examples}
 Suppose that $\alpha \le 0$ and $\ord_{\partial_1}(v_0)\le 0$  (recall that $v_0$ is the summand of $M_{00}$).  
Let $S(0) = 1+\sum_{\gamma =1}^{\infty}w_{\gamma}M_{00}^{-\gamma}$, where $\ord_{\partial_1}(w_{\gamma})\le -\alpha\gamma$. 
 
Then  the 
system $(SW)_{\alpha}$ with the initial condition $S(0)$ has a solution $S(t)\in \Nu_R$. 
 
 The solution $S(t)\in \Nu_R$ corresponds to a non-trivial solution of the 
system $(KP)_{\alpha}$ for sufficiently general $S(0)$ (see lemma \ref{admissible}).

\end{theo}

{\bf Proof.} (i) The idea of the proof is to look at the proof of theorem \ref{birkhoff} more carefully. Since the proof is very explicit, it is possible to check if the operator $S(t)$ constructed there belong to $\Nu_R$. The only difference with the proof of theorem \ref{birkhoff} is that we have to work with operators $U,S$ written as series in $L_{00}^{-1}, M_{00}^{-1}$ from the beginning of the proof.

So let $\exp (\sum_{i\in\sdz,j\in\sdz_+, i\le\alpha j}t_{ij}L_{00}^iM_{00}^j)S(0)^{-1}=U=\sum_{\beta\in \sdz}u_{\beta}M_{00}^{\beta}$ and 
$S(=S(t))=1+\sum_{\gamma =1}^{\infty}s_{\gamma}M_{00}^{-\gamma}$, where 
$u_{\beta},s_{\gamma}\in \widehat{E}_{\bar{R}}$. Now, to proceed the proof of theorem \ref{birkhoff}, we need a formula for the commutator $[M_{00}^k,a]$, $a\in \widehat{E}_{\bar{R}}$ for arbitrary $k,a$. Such a formula is in general too complicated, but we will need only some general properties of it, and we can use the notation and some basic ideas from \cite{Zh2}. Let's recall here some definition and lemmas from \cite{Zh2}. 

\begin{defin}
\label{maps}
Let us define linear maps ${}_m\delta_i: \widehat{E}_{\bar{R}}\rightarrow \widehat{E}_{\bar{R}}$, 
$m\in\dz$,
$i\in \dn$ as follows.
$$
M_{00}^maM_{00}^{-m}= {}_m\delta_0({a})+{}_m\delta_1({a})M_{00}^{-1}+
{}_m\delta_2({a})M_{00}^{-2}+\ldots ,\mbox{\quad}
a\in  \widehat{E}_{\bar{R}}.
$$

If $m= 0$, put ${}_m\delta_i= 0$. 
\end{defin}

Immediately from the definition follows

\begin{lemma}
\label{triviall}
In the situation of definition \ref{maps}
we have

(i) for $|m|>1$
$$
{}_m\delta_i(a)={}_{sign(m)}\delta_0 
({}_{sign(m)(|m|-1)}\delta_i(a))+
{}_{sign(m)}\delta_i({}_{sign(m)(|m|-1)}\delta_0(a))+
$$
$$
\sum_{j=1}^{i-1}{}_{sign(m)}\delta_j({}_{sign(m)(|m|-1)}\delta_{i-j}(a)),
$$
where $sign(m)=m/|m|$;

(ii) for any $m\ne 0$
$$
{}_{-m}\delta_0 ({}_{m}\delta_i)+
{}_{-m}\delta_i({}_{m}\delta_0)+
\sum_{j=1}^{i-1}{}_{-m}\delta_j({}_{m}\delta_{i-j})=0
$$
\end{lemma}





Suppose that the element $a$ from the definition belong to $E_{\bar R}$. Then,  
by formula (\ref{adm45}) from the proof of lemma \ref{admissible} (with $s_q$ replaced by $a$),  we can conclude that 
${}_m{\delta_{i}}(a)\in E_{\bar R}$ for all $i,m$, 
$\ord_{\partial_1}({}_m{\delta_{0}}(a))= \ord_{\partial_1}(a)$ for all $m$, and $\ord_{\partial_1}({}_1{\delta_{1}}(a))\le \ord_{\partial_1}(a)+\ord_{\partial_1}(v_0)$. Therefore, by lemma \ref{triviall}, we obtain $\ord_{\partial_1}({}_m{\delta_{1}}(a))\le \ord_{\partial_1}(a)+\ord_{\partial_1}(v_0)$ for all $m$, and, more generally, 
\begin{equation}
\label{nervo}
\ord_{\partial_1}({}_m{\delta_{i}}(a))\le \ord_{\partial_1}(a)+i\ord_{\partial_1}(v_0)
\end{equation}
for all $i\ge 1$.

Now we have 
\begin{multline*}
SU=(1+\sum_{\gamma =1}^{\infty}s_{\gamma}M_{00}^{-\gamma})(\sum_{\beta\in\sdz}u_{\beta}M_{00}^{\beta}) \\
=\sum_{\beta\in\sdz}u_{\beta}M_{00}^{\beta}+\sum_{\gamma =1}^{\infty}\sum_{\beta\in\sdz}\sum_{i=0}^{\infty}s_{\gamma}{}_{-\gamma}\delta_i(u_{\beta})M_{00}^{-\gamma +\beta -i}\\
=\sum_{\beta\in\sdz}u_{\beta}M_{00}^{\beta}+\sum_{\delta\in\sdz}(\sum_{\gamma =1}^{\infty}\sum_{i=0}^{\infty}s_{\gamma}{}_{-\gamma}\delta_i(u_{\delta +\gamma +i}))M_{00}^{\delta}.
\end{multline*}

Now the equation we have to solve is a system of algebraic equations
\begin{equation}
\label{3.8}
u_{-\beta}+\sum_{\gamma =1}^{\infty}\sum_{i=0}^{\infty}
s_{\gamma}{}_{-\gamma}\delta_i(u_{\gamma -\beta +i})=0 \mbox{\quad for $\beta =1,2,3,\ldots$}
\end{equation} 
Define
$$
\bfu =(u_{-1},u_{-2},u_{-3},\ldots ), \mbox{\quad } \bfs =(s_1,s_2,s_3, \ldots )
$$
and 
\begin{equation}
\label{Matr}
M=\left [
\sum_{i=0}^{\infty}{}_{-\gamma}\delta_i(u_{\gamma -\beta +i})
\right ]_{\gamma ,\beta =1,2,3, \ldots},
\end{equation}
where $M$ is a square matrix of infinite size with coefficients in $\widehat{E}_{\bar{R}}$. The solution $\bfs$ is given by $\bfs =-\bfu M^{-1}$.

Let $Q=1-M=[a_{\mu\nu}]_{\mu ,\nu=1,2,3,\ldots}$ and $Q^n=[a_{n,\mu\nu}]_{\mu ,\nu =1,2,3,\ldots }$. Since $a_{\mu\nu}=\delta_{\mu\nu}-\sum_{i=0}^{\infty}{}_{-\mu}\delta_i(u_{\mu -\nu +i})$, we have 
$$
 (a_{\mu\mu})|_{t=0}=1-(\sum_{i=0}^{\infty}{}_{-\mu}\delta_i(u_{\mu -\nu +i}))|_{t=0}=0,
$$
because of definition of $\widehat{E}_R^{\times}$, the property $\ord_{\partial_1}({}_m{\delta_{0}}(a))= \ord_{\partial_1}(a)$ and the obvious fact that the maps ${}_m{\delta_{i}}$ are $k[\ldots ,t_{ij},\ldots ]$-linear. Similarly, if $\mu >\nu$, then 
$$
\pi (a_{\mu\nu})=\pi (-\sum_{i=0}^{\infty}{}_{-\mu}\delta_i(u_{\mu -\nu +i}))=0.
$$
Because of the growth order condition for $u_{\nu}$, we can find a positive real number $J$ such that 
\begin{equation}
\label{3.10000}
v_2(a_{\mu\nu})\ge J(\mu -\nu ) \mbox{\quad for all $\mu -\nu\ge 0$}.
\end{equation}
as before.
 
{\it Claim.} i) For every $n\ge 1$ we have $v_2(a_{n,\mu \nu})\ge J(\mu -\nu )$ if $\mu -\nu \ge 0$. 

ii) The function $F(M,N,a_{n,\mu \nu})$ satisfy the following property: for any given $M,N,\mu ,\nu$ there exists a natural number $T(M,N,\mu ,\nu )$ such that $F(M,N,a_{n,\mu \nu})=0$ for all $n>T(M,N,\mu ,\nu )$.

iii) $\ord_{\partial_1}(a_{n,\mu\nu})\le \alpha (\mu -\nu )$. 

{\it Proof of the claim.} To prove i), ii) we can repeat the proof of the claim in theorem \ref{birkhoff}. Let us prove iii). 

Let $U_0=\exp (\sum_{i\in\sdz,j\in\sdz_+, i\le\alpha j}t_{ij}L_{00}^iM_{00}^j)=1+ \sum_{\beta =1}^{\infty}u_{\beta}^0M_{00}^{\beta}$ and $S(0)^{-1}=1+\sum_{k=1}^{\infty}s_k^0M_{00}^{-k}$. 
Note that, because of (\ref{nervo}) and the condition $\ord_{\partial_1}(v_0)\le 0$, the coefficients $s_k^0$ satisfy the same condition as the coefficients $w_k$ of the operator $S(0)$, i.e. $\ord_{\partial_1}(s_{k}^0)\le -\alpha k$. 
Now 
\begin{multline*}
U=U_0S(0)^{-1}=(1+\sum_{\beta =1}^{\infty}u_{\beta}^0M_{00}^{\beta})(1+\sum_{k=1}^{\infty}s_k^0M_{00}^{-k}) \\
=1+\sum_{\beta =1}^{\infty}u_{\beta}^0M_{00}^{\beta}+\sum_{k=1}^{\infty}s_k^0M_{00}^{-k}+\sum_{\beta =1}^{\infty}\sum_{k=1}^{\infty}\sum_{i=0}^{\infty}u_{\beta}^0{}_{\beta}\delta_i(s_k^{0})M_{00}^{\beta -k-i} \\
=1+\sum_{\beta =1}^{\infty}u_{\beta}^0M_{00}^{\beta}+\sum_{k=1}^{\infty}s_k^0M_{00}^{-k}+\sum_{\gamma \in\sdz }(\sum_{\beta =1}^{\infty}\sum_{i=0}^{\infty}u_{\beta}^0{}_{\beta}\delta_i(s_{\beta -i-\gamma}^{0}))M_{00}^{\gamma},
\end{multline*}
where ${}_{\beta}\delta_i(s_{\beta -i-\gamma}^{0})=0$ if $\beta -i-\gamma \le 0$. 
Since $\ord_{\partial_1}(v_0)\le 0$, we have by (\ref{nervo}) 
$$
\ord_{\partial_1}(u_{\beta}^0{}_{\beta}\delta_i(s_{\beta -i-\gamma}^{0}))\le\ord_{\partial_1}(u_{\beta}^0)+
\ord_{\partial_1}({}_{\beta}\delta_i(s_{\beta -i-\gamma}^{0}))\le 
$$
$$
\ord_{\partial_1}(u_{\beta}^0) + 
\ord_{\partial_1}(s_{\beta -i-\gamma}^{0})\le \alpha \beta + \alpha (-\beta +i+\gamma )\le \alpha\gamma ,
$$
where from we get $\ord_{\partial_1}(u_{\gamma})\le \alpha \gamma$. Applying again (\ref{nervo}),  
we get 
$$
\ord_{\partial_1}(a_{1,\mu\nu})\le \ord_{\partial_1}(u_{\mu -\nu})\le \alpha (\mu -\nu ).
$$
Assume now that iii) holds for some $n\ge 1$. Since  
$
a_{n+1,\mu\nu}=\sum_{l=1}^{\infty}a_{n,\mu l}a_{l\nu},
$
we get 
$$
\ord_{\partial_1}(a_{n+1,\mu\nu})\le \ord_{\partial_1}(a_{n,\mu l}a_{l\nu})\le \alpha (\mu -l) +\alpha (l-\nu )=\alpha (\mu - \nu ).
$$ 
This completes the proof of the claim.

By this claim we can conclude that $\sum_{n=0}^{\infty}a_{n,\mu\nu}\in E_{\bar R}$ is well defined for all $\mu ,\nu=1,2,3,\ldots $. Therefore, 
$$
M^{-1}=\sum_{n=0}^{\infty}(1-M)^n=[\sum_{n=0}^{\infty}a_{n,\mu\nu}]_{\mu ,\nu}
$$
is well defined. Let $M^{-1}=[b_{\mu\nu}]_{\mu ,\nu =1,2,3,\ldots }$, namely $b_{\mu\nu}=\sum_{n=0}^{\infty}a_{n,\mu\nu}$. Then 
$$
s_{\nu}=-\sum_{\mu =1}^{\infty}u_{-\mu}b_{\mu\nu}\in E_{\bar R}
$$
is a well-defined element with $\ord_{\partial_1}(s_{\nu})\le \ord_{\partial_1}(u_{-\mu})+\ord_{\partial_1}(b_{\mu\nu})\le -\alpha \nu$. Thus we have established the existence of $S=S(t)\in 1+E_{R-}$ such that $SU\in \widehat{D}_R$. 

At last, by lemma \ref{admissible}, the operator $S(0)$ of a sufficiently general type is not admissible, that is the solution $S(t)$ corresponds to a non-trivial solutions of the system $(KP)_{\alpha}$. \\
The theorem is proved.\\
$\Box$

\section{Isospectral deformations and the modified KP systems}

There is a natural question: can we obtain the Parshin system or the modified systems as defining equations of all isospectral deformations of some differential operators in two variables? We will show below that it is true for a pair of monic commuting differential operators. 

The notion of  isospectral deformations can be introduced in the same way as in \cite{Mu3}, \S 4. Consider a family 
$$
\{P(t), t\in M\}
$$
of operators, where the parameter space $M$ is an open domain of $\dc^N$ and $P(t)=(P_1(t), P_2(t))$, $P_i\in A[t]((\partial_1^{-1}))[\partial_2]=\tilde{D}\subset D_R$ is a pair of  monic commuting "differential" operators  depending on  $t=(t_1,\ldots ,t_N)\in M\subset \dc^N$ analytically. As a specialization we can take, for example, $A=\dc [[x_1,x_2]]$, $\partial_1=\partial /\partial x_1$, $\partial_2=\partial /\partial x_2$. For convenience, further we will work with this specialization. 
\begin{defin}
\label{4.0}
We say $\{P(t), t\in M\}$ is a family of isospectral deformations if there exist  
"differential operators" $Q_1(t), Q_2(t),\ldots , Q_N(t)\in \tilde{D}$ depending on the parameter 
$t\in M$ analytically such that the following system of equations has a nontrivial solution $\psi (A,t;\lambda )$  for every eigenvalue $\lambda =(\lambda_1,\lambda_2)\in \dc^2$ of $P(t)$: 
\begin{equation}
\label{4.4.1}
\left \{
\begin{array}{c}

P(t)\psi (A,t;\lambda )=\lambda \psi (A,t;\lambda )\\
\frac{\partial}{\partial t_1}\psi (A,t;\lambda )=Q_1(t)\psi (A,t;\lambda )\\
\ldots \\
\frac{\partial}{\partial t_N}\psi (A,t;\lambda )=Q_N(t)\psi (A,t;\lambda )\\
\end{array}
\right .
\end{equation}

\end{defin}

The point here is that the eigenvalue $\lambda$ in the first equation does not depend on the parameter $t$, i.e., it is preserved. 
Repeating the arguments from \cite{Mu3}, \S 4, we obtain the compatibility conditions of the system (\ref{4.4.1}):
$$
0=\frac{\partial}{\partial t_i}(P(t)\psi (x,t;\lambda )-\lambda \psi (x,t;\lambda ))= (\frac{\partial}{\partial t_i}P(t) - [Q_i(t), P(t)])\psi (x,t;\lambda ),
$$ 
where $x=(x_1,x_2)$. 
For every fixed $t\in M$, the eigenfunctions $ \psi (x,t;\lambda )$ are linearly independent for distinct eigenvalues $\lambda\in \dc$. Since $\frac{\partial}{\partial t_i}P_j(t) - [Q_i(t), P_j(t)]$, $j=1,2$ are  pseudo-differential operators of finite order in $\partial_2$, they have mostly a countable (topological) basis of independent solutions. Therefore, by cardinality reason, 
\begin{equation}
\label{4.2}
\frac{\partial}{\partial t_i}P(t)=[Q_i(t), P(t)]
\end{equation}

Similarly, the condition  $\frac{\partial}{\partial t_i}\frac{\partial}{\partial t_j}\psi =\frac{\partial}{\partial t_j}\frac{\partial}{\partial t_i}\psi $ gives 
\begin{equation}
\label{4.4.3}
\frac{\partial}{\partial t_i}Q_j(t)-\frac{\partial}{\partial t_j}Q_i=[Q_i(t), Q_j(t)].
\end{equation}

The system of equations (\ref{4.2}) and (\ref{4.4.3}) is equivalent to the condition that equation (\ref{4.4.1}) has a nontrivial solution for every $\lambda\in \dc^2$. Therefore, finding a family $P(t)$ of isospectral deformations of a given pair $P(0)$ is equivalent to finding a solution of the Lax equation (\ref{4.2}) for differential operators $Q_i(t)$ satisfying (\ref{4.4.3}) together with the initial condition $P(t)|_{t=0}=P(0)$. 

Without loss of generality assume that $\ord_{\partial_2}(P_1)\ge \ord_{\partial_2}(P_2)$. Let 
$$(\ord_{\partial_1}(P_1\partial_2^{-\ord_{\partial_2}(P_1)})_+, \ord_{\partial_2}(P_1))=(p_1,q_1)$$ 
and 
$$(\ord_{\partial_1}(P_2\partial_2^{-\ord_{\partial_2}(P_2)})_+, \ord_{\partial_2}(P_2))=(p_2,q_2).$$ 
For each pseudo-differential operator we will call such a pair of integers {\it the full order}. 

\begin{lemma}
\label{equivvv}
Suppose $(p_1,q_1)\ne d(p_2/l,q_2/l)$ for any $d\in\dz$, where $l=gcd(p_2,q_2)$. Then equation (\ref{4.2}) is equivalent to the equation 
\begin{equation}
\label{83}
\frac{\partial}{\partial t_i}L(t)=[Q_i(t), L(t)], 
\end{equation}
where $L=(L_1,L_2)$, 
$$L_1=u_0+u_{1}\partial_2^{-1}+\ldots ,$$ 
$u_i\in A[t]((\partial_1^{-1}))$, $\ord_{\partial_1}(u_0)=1$, $u_0$ is monic, 
$$
L_2=v_{-1}\partial_2+v_0+v_{1}\partial_2^{-1}+\ldots ,
$$
$u_i\in A[t]((\partial_1^{-1}))$, $\ord_{\partial_1}(v_{-1})=0$, $v_{-1}$ is monic.

\end{lemma}

{\bf Proof.} Consider the operator $P'_1:=P_1^{q_2/(l\cdot gcd(q_1,q_2/l))}P_2^{-(1/l)q_1/gcd(q_1,q_2/l)}$. Let the full order of $P'_1$ be equal to $(k,0)$, $k\in\dz$.  

If $k=0$, this means that $p_1q_2/(l\cdot gcd(q_1,q_2/l))=p_2q_1/(l\cdot gcd(q_1,q_2/l))$, where from $p_2/l$ is divisible by $q_2/(l\cdot gcd(q_1,q_2/l))$ and $p_1$ is divisible by $q_1/gcd(q_1,q_2/l)$. Since $(p_2/l, q_2/l)=1$, we have therefore $gcd(q_1,q_2/l)=q_2/l$ and $(p_1,q_1)=q_1l/q_2(p_2/l, q_2/l)$, a contradiction.  

So, $k\ne 0$, and we put $L_1:={P'_1}^{1/k}$. Since $P_1,P_2$ are monic operators, such a root exists. Then $L_2:=(P_2L_1^{-p_2})^{1/q_2}$. Clearly, equation (\ref{83}) implies equation (\ref{4.2}). Let's prove the converse. 

Since $\frac{\partial}{\partial t_i}$ and $[Q_i(t), . ]$ are derivations, we have 
$$
0=(\frac{\partial}{\partial t_i} -[Q_i(t), . ])P_2(t)=\sum_{k=0}^{l-1}P_2(t)^{k/l}(\frac{\partial}{\partial t_i}P_2(t)^{1/l} -[Q_i(t), P_2(t)^{1/l}])P_2(t)^{(l-1-k)/l}
$$
Since $P_2(t)^{1/l}$ is monic, the last equality implies that $\frac{\partial}{\partial t_i}P_2(t)^{1/l} -[Q_i(t), P_2(t)^{1/l}]=0$. Therefore, equation (\ref{4.2}) is equivalent to the equation 
$$
\frac{\partial}{\partial t_i}P_2(t)^{1/l} =[Q_i(t), P_2(t)^{1/l}],\mbox{\quad} \frac{\partial}{\partial t_i}P_1(t)=[Q_i(t), P_1(t)]
$$
Continuing this line of reasoning, we obtain the equivalence of the equation above with equation (\ref{83}).\\
$\Box$

{\bf Remark.} The condition on operators $P_1,P_2$ in lemma appears as an analog of condition of ellipticity for commuting differential operators in one variable (see \cite{Mu2}, sect.5). In one-dimensional situation the existence of a monic operator in a ring of commuting operators implies the moniqueness of all commuting operators up to a constant. In our case the existence of two monic operators is not enough for having an analogous property in the ring of commuting operators. The extra condition from lemma imply this property as we will see from lemma below. Obviously, the operators from lemma are also algebraically independent. In another paper we are going to show that rings of commuting operators belonging to some subspaces of the ring of "differential" operators that have two such operators correspond to certain geometric data and vice versa. Some examples of such rings of commuting variables can be obtained as certain images of rings considered in \cite{ZhO}, theorem 1, b) and remark 3. They appear as images of a generalized Krichever map constructed in \cite{Pa3}, \cite{O}. Other examples should (conjecturally) come from the examples considered in \cite{Et}, \cite{Ga}, where the rings of commuting differential operators containing completely integrable operators of dimension two appeared. See also a discussion below. 

\smallskip

The left hand side of 
equation (\ref{83}) is a pair of operators of orders (in $\partial_2$) at most $(0,1)$. Therefore, the operator $Q_i(t)$ must satisfy 
\begin{equation}
\label{chtoto}
\ord_{\partial_2}([Q_i(t), L_1(t)])\le 0,\mbox{\quad} \ord_{\partial_2}([Q_i(t), L_2(t)])\le 1.
\end{equation}
For simplicity, we can assume that operators $L_1,L_2$ (so, $P_1,P_2$) are normalized. By this we mean the operators obtained by conjugation by some invertible operator $\bar S\in \tilde{D}^{\times}$ as in theorem \ref{Parshin}, iii) or in the proof of lemma \ref{admissible}. In this case canonically defined by $L_1,L_2$ elements $L_{00},M_{00}$ have some special form. Let's assume also that $L_{1+}, L_{2+}$ do not depend on times and that $v_{-1}=1$ (recall that $M_{00}=v_{-1}\partial_2+v_0$). In this case (\ref{chtoto}) becomes 
\begin{equation}
\label{chtoto1}
\ord_{\partial_2}([Q_i(t), L_1(t)])< 0,\mbox{\quad} \ord_{\partial_2}([Q_i(t), L_2(t)])< 0.
\end{equation}

\begin{lemma}
\label{12}
Let $L=(L_1,L_2)$, $L_1,L_2\in A((\partial_1^{-1}))((\partial_2^{-1}))$ be arbitrary monic operators with $\ord_{\partial_2}(L_1)=0$, $\ord_{\partial_2}(L_2)=1$, $\ord_{\partial_1}(L_{1+})=1$, $\ord_{\partial_1}((L_2\partial_2^{-1})_+)=0$. Then 
$$
F_L=\{Q\in \tilde{D}| \ord_{\partial_2}([Q_i(t), L_1(t)])< 0,\mbox{\quad} \ord_{\partial_2}([Q_i(t), L_2(t)])< 0\}
$$
coincides with the $\dc$-linear space (topologically) generated by the operators $(L_1^iL_2^j)_+$, $i\in\dz$, $j\in\dz_+$. 
\end{lemma} 

{\bf Proof.} Obviously, the operators $(L_1^iL_2^j)_+$ belong to $F_L$. Conversely, let $Q\in F_L$ be an element of full order $(m_1,m_2)$. We can represent $Q$ as $S^{-1}Q'S$, where $S$ is the operator from theorem \ref{Parshin} for $L_1,L_2$, and $Q'$ is a series in $M_{00}^{-1}$ with coefficients represented as series in $L_{00}^{-1}$ ($L_{00}=\partial_1$ if $L$ is normalized as in the proof of lemma \ref{admissible}). The condition $\ord_{\partial_2}([Q_i(t), L_1(t)])< 0$ then implies that coefficients of all non-negative powers of $M_{00}$ of this series (which are series in  $L_{00}^{-1}$) commute with $L_{00}$. Since we assumed that $L$ is normalized, $v_0$ also commutes with $L_{00}$, and therefore also with these coefficients. The second condition $\ord_{\partial_2}([Q_i(t), L_2(t)])< 0$ then implies that coefficients of all non-negative powers of $M_{00}$ commute with $\partial_2$, and therefore they are constant. 

The last fact means that the leading coefficient of the leading coefficient of $Q$ is constant, say $c\in\dc$. Since $(L_1^{m_1}L_2^{m_2})_+$ is monic, the linear combination $Q-c(L_1^{m_1}L_2^{m_2})_+$ has order less than $(m_1,m_2)$. Since $(Q-c(L_1^{m_1}L_2^{m_2})_+)$ satisfy (\ref{chtoto1}), the lemma follows by induction on $(m_1,m_2)$. 

Note that we have got a representation of $Q$ in an {\it infinite} sum, $Q=\sum_{j=0}^{m_2}\sum_{i=-\infty}^{i_j} c_{ij}(L_1^iL_2^j)_+$. This means that $(L_1^iL_2^j)_+$ are topological generators of the vector space $F_L$, where the topology comes from the topology of a two-dimensional locla field $\dc ((L_1^{-1}))((L_2^{-2}))$. Indeed, 
$$
\sum_{j=0}^{m_2}\sum_{i=-\infty}^{i_j} c_{ij}(L_1^iL_2^j)_+=(\sum_{j=0}^{m_2}\sum_{i=-\infty}^{i_j} c_{ij}L_1^iL_2^j)_+=
(S^{-1}(\sum_{j=0}^{m_2}\sum_{i=-\infty}^{i_j} c_{ij}L_{00}^iM_{00}^j)S)_+.
$$
The lemma is proved.\\
$\Box$

By proposition \ref{svyaznost'} the equation 
$$
\frac{\partial}{\partial t_{ij}}L(t)=[(L_1(t)^{i}L_2(t)^{j})_+,L(t)], \mbox{\quad} i\in\dz ,j\in \dz_+
$$
imply equation (\ref{4.4.3}). Thus it is the master equation for the largest possible family of isospectral deformations of a pair of "differential" operators satisfying the condition of lemma \ref{equivvv} and assumptions above. 

Note that the original Parshin's system (\ref{Par}) is a "half" of the system above. Therefore, as we have shown in section 2.2, it has only trivial solutions. This means that in general it is not possible to find a universal family. Nevertheless, it is possible to find some kind of restricted universal family of isospectral deformations, where we mean that the operators $Q_i$ should belong not to the whole space of differential operators, but to some linear subspace generated by operators with some restrictions on the full order. Let's clarify the last assertion.  

For any pair of operators $P_1,P_2\in D_{R}$ satisfying the condition of lemma \ref{equivvv} the operators $L_1,L_2$ and the invertible operator $S\in \widehat{\Nu}_R$ are defined so that $L_1=S^{-1}L_{00}S$, $L_2=S^{-1}M_{00}S$. Then $SP_iS^{-1}=NP_i$, $i=1,2$, where $NP_i$ are monomials in $L_{00},M_{00}$. We will call this pair of operators {\it a normal form} of the pair of operators $P_1,P_2$. 

For any function $\alpha :\dz_+ \rightarrow \dr$, an operator $W\in \widehat{\Nu}_R$ and monic operators $L_{00}\in E_{\bar R}$, $M_{00}=v_{-1}\partial_2 +v_0$, $v_i\in E_{\bar R}$ let $D_R^{\alpha ,W, L_{00}, M_{00}}$ be a vector space defined as follows
$$
D_R^{\alpha ,W,L_{00},M_{00}}=\langle (W^{-1}QW)_+ | Q=\sum_{k=0}^Nq_kM_{00}^k\in \bar{R}((L_{00}^{-1}))[M_{00}], \mbox{\quad } \ord_{\partial_1}(q_k)\le \alpha (k)\rangle .
$$ 
Obviously, $D_R=\cup_{\alpha}D_R^{\alpha ,W,L_{00},M_{00}}$ for any fixed $W, L_{00},M_{00}$, where the union is taken over all possible functions $\alpha$. Of course, in general the space $D_R^{\alpha ,W,L_{00},M_{00}}$ is far from to be a ring. But in some partial cases, for example for linear functions and operators $L_{00},M_{00}$ considered in section 5, $W=1$, this space has a natural ring structure, as one can easily check.

Now, let $P_1(0),P_2(0)\in A((\partial_1^{-1}))[\partial_2]$ be any pair of operators  satisfying the condition of lemma \ref{equivvv}. Let's fix the operators $L_{00},M_{00}$ defined by this pair and some function $\alpha$ such that $P_i(0)\in D_R^{\alpha ,S(0),L_{00},M_{00}}$. Obviously, such a function exists. Moreover, since the normal form of $P(0)$ consists of monomials, we can assume that $\alpha (0)\le 0$. We will say that operators belonging to $D_R^{\alpha ,\ldots }$ are {\it $\alpha$-differential operators}. All said above brings the following definition. 

\begin{defin}
\label{4.0.'}
We say that a family $\{P(t), t\in M\}$ of pairs of monic commuting $\alpha$-differential operators, whose full orders are constant and  satisfy the condition of lemma \ref{equivvv}, $P(t)=(P_1(t), P_2(t)), P_i(t)\in \tilde{D}\cap D_R^{\alpha ,S(t),L_{00},M_{00}}$ is a family of isospectral deformations if there exist  
$\alpha$-differential operators $Q_1(t), Q_2(t),\ldots , Q_N(t)\in \tilde{D}\cap D_R^{\alpha ,S(t),L_{00},M_{00}}$ depending on the parameter 
$t\in M$ analytically such that the following system of equations has a nontrivial solution $\psi (A,t;\lambda )$  for every eigenvalue $\lambda =(\lambda_1,\lambda_2)\in \dc^2$ of $P(t)$: 
\begin{equation}
\label{4.4.1'}
\left \{
\begin{array}{c}
P(t)\psi (A,t;\lambda )=\lambda \psi (A,t;\lambda )\\
\frac{\partial}{\partial t_1}\psi (A,t;\lambda )=Q_1(t)\psi (A,t;\lambda )\\
\ldots \\
\frac{\partial}{\partial t_N}\psi (A,t;\lambda )=Q_N(t)\psi (A,t;\lambda )\\
\end{array}
\right .
\end{equation}

\end{defin}

Repeating all arguments after definition \ref{4.4.1} with $\tilde{D}$ replaced by $\tilde{D}\cap D_R^{\alpha ,S(t),L_{00},M_{00}}$ and indices $i\in\dz$, $j\in\dz_+$ replaced by $i\in\dz$, $j\in\dz_+$, $i\le \alpha (j)$ in lemma \ref{12}, we come to a conclusion that the equation $(KP)_{\alpha}$ (see sect. 3.1) is the master equation for the largest possible family of isospectral deformations of a pair of $\alpha$-differential operators satisfying the condition of lemma \ref{equivvv}. 

As we have shown in sections 3,4, this equation is uniquely solvable. By the same arguments as in \cite{Mu3}, section 4 (equation (4.16)), we obtain 
$$
P(t)=S(t)NP(0)S(t)^{-1}=Y(t)P(0)Y(t)^{-1},
$$
where $S(t)$ is the solution of the corresponding modified Sato-Wilson system, and $Y(t)$ is the  operator coming from theorem \ref{birkhoff}. This shows that the pair of operators $P_i(t)\in D_R^{\alpha ,S(t),L_{00},M_{00}}$, because they have the same normal form as $P(0)$ and because $Y(t)$ has no negative order terms ($P_i(t)\in \tilde{D}\cap D_R^{\alpha ,S(t),L_{00},M_{00}}$ if $S(0)$ looks like in theorem \ref{examples}). 

So, the systems $(KP)_{\alpha}$ give for certain pairs $P$ of $\alpha$-differential operators a universal family of isospectral deformations.

\end{document}